\documentclass[12pt]{article}

\usepackage{paralist}
\usepackage{wrapfig}
\usepackage{graphicx}
\usepackage[none]{hyphenat}
\usepackage[english]{babel}
\usepackage[utf8]{inputenc}
\usepackage[T1]{fontenc}
\usepackage{tikz-cd,pgfplots}
\usepackage{epsfig}

\usepackage{epstopdf}

\usepackage{parskip}

\usepackage{authblk}

\usetikzlibrary{calc, patterns,arrows.meta, arrows, shapes.geometric}

\usepackage{graphicx,amsmath,amssymb,url,enumerate,mathrsfs,epsfig,color}

\usetikzlibrary{decorations.text}
\usetikzlibrary{decorations.markings}
\usetikzlibrary{calc, patterns,arrows, shapes.geometric}

\usepackage{caption}

\usepackage{csquotes}
\usepackage{mathtools,amssymb,amsfonts,amsthm}

\usepackage[breaklinks,colorlinks=true]{hyperref}
\hypersetup{urlcolor=blue, citecolor=red}
\usepackage[capitalise]{cleveref} 

\theoremstyle{plain}

\newtheorem*{theorem*}{Theorem}

\newtheorem*{lemma*}{Lemma}
\newtheorem{proposition}{Proposition}
\newtheorem*{proposition*}{Proposition}
\newtheorem{corollary}{Corollary}
\newtheorem*{corollary*}{Corollary}
\newtheorem{remark}{Remark}

\theoremstyle{definition}
\newtheorem{definition}{Definition}
\newtheorem*{definition*}{Definition}
\newtheorem{example}{Example}
\newtheorem*{example*}{Example}

\crefname{theorem}{Theorem}{Theorems}
\Crefname{theorem}{Theorem}{Theorems}
\crefname{lemma}{Lemma}{Lemmas}
\Crefname{lemma}{Lemma}{Lemmas}
\crefname{proposition}{Proposition}{Propositions}
\Crefname{Prop}{Proposition}{Propositions}
\crefname{corollary}{Corollary}{Corollaries}
\Crefname{corollary}{Corollary}{Corollaries}
\crefname{definition}{Definition}{Definitions}
\Crefname{definition}{Definition}{Definitions}
\crefname{example}{Example}{Examples}
\Crefname{example}{Example}{Examples}
\crefname{theorem}{Theorem}{Theorems}

\newtheorem*{exercise*}{Exercise}
\crefname{exercise}{exercise}{exercises}
\Crefname{exercise}{Exercise}{Exercises}

\newcommand{\hor}{\mathrlap{\perp}\square}
\newcommand{\ram}{\raisebox{0.22ex}{$-$}}
\newcommand{\ver}{{\mathrlap{\ram}}\square}

\hypersetup{urlcolor=blue, citecolor=red}

\title{Double groupoids of composites: applications to uniformity}

\author{
\begin{center}

V{\'\i}ctor M. Jim\'enez\footnote{E-mail: \href{mailto:victor.jimenez@mat.uned.es}{victor.jimenez@mat.uned.es}}
\\  Universidad Nacional de Educación a Distancia (UNED), \\
Departamento de Matemáticas Fundamentales. \\
Calle de Juan del Rosal 10, 28040, Madrid, Spain

\bigskip

Manuel de Le\'on\footnote{E-mail: \href{mailto:mdeleon@icmat.es}{mdeleon@icmat.es}}
\\ Instituto de Ciencias Matem\'aticas, Campus Cantoblanco \\
 Consejo Superior de Investigaciones Cient\'ificas
 \\
C/ Nicol\'as Cabrera, 13--15, 28049, Madrid, Spain
\\
and
\\
Real Academia de Ciencias de España.
\\
C/ Valverde, 22, 28004 Madrid, Spain.

Marcelo Epstein \footnote{E-mail: \href{mailto:mepstein@ucalgary.ca}{mepstein@ucalgary.ca}}
\\  University of Calgary\\
Calgary, Alberta T2N1N4
Canada.

\bigskip
\end{center}
}

\begin{document}
\maketitle

\begin{abstract}
In this paper we present a geometrical framework to study the uniformity of a composite material by means of a material double groupoid. The notions of vertical and horizontal uniformity are introduced, as well as other weaker ones that allows us to study other possible notions of more general uniformity.
\end{abstract}
\tableofcontents
\section{Introduction}
\sloppy

In continuum physics, the theory of material uniformity addresses the question of comparing the properties of pairs of points in a material substrate, assumed to be continuous. Uniformity is declared when the result of this investigation is that, according to a priori established criteria, those properties are identical for every pair (see \cite{WNOLL,CTRUE,CCWAN}, and \cite{VMMDME,EPSBOOK} for a modern description using the language of differential geometry).

Even if the substrate turns out to be uniform, however, the question can be raised as to whether there exists a configuration in which all points are simultaneously in the same state. A positive answer to this question gives rise to the notion of homogeneity. Lack of homogeneity of a uniform body can be related to various physical theories of continuous distributions of dislocations and other defects, as well as to the presence of residual stresses.  Although the study of the homogeneity of composite materials is of great relevance, it will not be addressed in this paper since we need to first understand in depth the concept of uniformity.

Indeed, there is the possibility of the coexistence of different kinds of properties in the same material substrate, for instance, in the case of material composites in which the constituents, although blended together, retain their individual original structure. Each of these components may be perfectly uniform in its own right, but the result will not be so in general.

In this paper we will focus on the study of the uniformity of composite materials, postponing the study of homogeneity to a later work, as we have mentioned above. In the case of uniformity we face a practically unexplored terrain, for which we adopt as methodology and instruments the concept of groupoid and, in particular, the double groupoid (see \cite{KMG} for a general reference of the theory of groupoids). This methodology has been very useful in previous works (see \cite{VMMDME,MATGROUPALG,MGEOEPS}). 

As a first approach, presented in \cite{eps01}, we will consider the scenario of a composite material, a solid mixture, or a material characterized by multiple independent constitutive structures. A straightforward approach to address such cases is to identify the symmetries shared by the individual components and determine what can be regarded as the intersection of their respective material groupoids. In other words, if we have a composite material made of two different materials with mechanical responses $W_1$ and $W_2$, namely
$$
W_1 = W_1(X, F) \; , \; 
W_2 = W_2(X, F) ,
$$
identically for all deformation gradients $F$, we obtain two material subgroupoids of the 1-jet groupoid of $B$,
$\Omega_1(B)$ and $\Omega_2(B)$, respectively.
So, it is natural to think that the composite is uniform if,
$$
\Omega_1(B) (X, Y) \cap \Omega_2(B) (X, Y) \not= \emptyset
$$
for any two points $X, Y \in B$. Here, $\Omega (X, Y)$ denotes the set of arrows from $X$ to $Y$ for an arbitrary groupoid $\Omega$. But, in any case, even if $B$ is uniform for the two material structures simultaneously, we can not guarantee that it is uniform as a composite.

Also, we can consider the intersection
$
\Omega_1(B) \cap \Omega_2(B) ,
$
which is a subgroupoid of $\Pi^1(B, B)$, and we can investigate under what conditions $\Omega_1(B) \cap \Omega_2(B)$ is a Lie subgroupoid assuming that both $\Omega_1(B)$ and $\Omega_2(B)$ are Lie subgroupoids. 
Notice that this assertion is true for two Lie sugbroups of a given Lie group, but it is not true in general in the context of Lie groupoids.\\

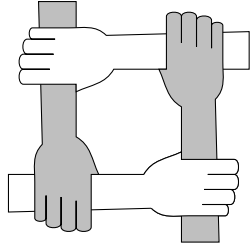
\begin{figure}[h!]
\begin{center}
\begin{tikzpicture} [scale=1.0]
\draw  (2.1,0.35)--(0.2,0.3)--(0.2,0.8) -- (2,0.8) to[out=60,in=205] (2.5,1.1)--(2.5,1)--(3.2,1) to [out=0, in=0] (3.2,0.8) -- (2.85,0.8) -- (3.2,0.8) to [out=0, in=0] (3.2,0.6)--(2.8,0.6)--(3.15,0.6) to [out=0, in=0] (3.15,0.4)--(2.78,0.4)--(3.08,0.4) to [out=0, in=0] (3.08,0.25)--(2.6,0.25) to [out=180,in=-45] (2.1,0.35) ;


\begin{scope} [rotate around={270:(1.80,1.5)},yshift=0]
\draw[fill=gray!50]  (2.1,0.35)--(0.2,0.3)--(0.2,0.8) -- (2,0.8) to[out=60,in=205] (2.5,1.1)--(2.5,1)--(3.2,1) to [out=0, in=0] (3.2,0.8) -- (2.85,0.8) -- (3.2,0.8) to [out=0, in=0] (3.2,0.6)--(2.8,0.6)--(3.15,0.6) to [out=0, in=0] (3.15,0.4)--(2.78,0.4)--(3.08,0.4) to [out=0, in=0] (3.08,0.25)--(2.6,0.25) to [out=180,in=-45] (2.1,0.35) ;
\end{scope}
\begin{scope} [rotate around={180:(1.80,1.5)},yshift=0]
\draw[fill=white]  (2.1,0.35)--(0.2,0.3)--(0.2,0.8) -- (2,0.8) to[out=60,in=205] (2.5,1.1)--(2.5,1)--(3.2,1) to [out=0, in=0] (3.2,0.8) -- (2.85,0.8) -- (3.2,0.8) to [out=0, in=0] (3.2,0.6)--(2.8,0.6)--(3.15,0.6) to [out=0, in=0] (3.15,0.4)--(2.78,0.4)--(3.08,0.4) to [out=0, in=0] (3.08,0.25)--(2.6,0.25) to [out=180,in=-45] (2.1,0.35) ;
\end{scope}
\begin{scope} [rotate around={90:(1.80,1.5)},yshift=0]
\draw[fill=gray!50]  (2.1,0.35)--(0.2,0.3)--(0.2,0.8) -- (2,0.8) to[out=60,in=205] (2.5,1.1)--(2.5,1)--(3.2,1) to [out=0, in=0] (3.2,0.8) -- (2.85,0.8) -- (3.2,0.8) to [out=0, in=0] (3.2,0.6)--(2.8,0.6)--(3.15,0.6) to [out=0, in=0] (3.15,0.4)--(2.78,0.4)--(3.08,0.4) to [out=0, in=0] (3.08,0.25)--(2.6,0.25) to [out=180,in=-45] (2.1,0.35) ;
\end{scope}
\draw[fill=white]  (2,0.8) to[out=60,in=205] (2.5,1.1)--(2.5,1)--(3.2,1) to [out=0, in=0] (3.2,0.8) -- (2.85,0.8) -- (3.2,0.8) to [out=0, in=0] (3.2,0.6)--(2.8,0.6)--(3.15,0.6) to [out=0, in=0] (3.15,0.4)--(2.78,0.4)--(3.08,0.4) to [out=0, in=0] (3.08,0.25)--(2.6,0.25) to [out=180,in=-45] (2.1,0.35) ;
\end{tikzpicture}
\end{center}
\caption{Artistic rendition of the building block of a double groupoid \cite{eps02}.}
\label{fig:hands}
\end{figure}

In a series of papers \cite{eps01,eps02,eps03}, a first approach was suggested for the study of composite materials using the notion of \textit{double groupoids}\footnote{Double categories and double groupoids were introduced in 1963 by Charles Ehresmann \cite{ehresmann}.}, an idea that seems very natural, as a unified superstructure which combines both material structures. The fundamental notion of a double groupoid can be interpreted as the structured interaction of two pairs of material isomorphisms, which combine to form a coherent mathematical entity, as vaguely suggested in Figure \ref{fig:hands}.

Because the aim of the present paper is to study the uniformity properties of a composite material, and since in simple materials that notion is linked to the transitivity of the associated material groupoid, it is clear that the methodology to be used relies on the different notions of transitivity of a double Lie groupoid. Indeed, the abstract structure of double groupoid allows us to introduce
several notions of transitivity: horizontal and vertical transitivity, strong and weak transitivity. All these notions are interpreted as different kinds of uniformity on the composite material. 

The paper is structured as follows. To be as self-contained as possible, we will recall the elementary notions of simple materials and their uniformity in section 2. Section 3 is devoted to introduce the notion of groupoid and double groupoid as well as their main properties. Although working with double groupoids is complicated due to the number of arrows at play, the use of commutative squares has allowed us to express many of the concepts in a visual and intuitive way. The different notions of transitivity in a double groupoid introduced in this last section are translated to the study of the uniformity of composite materials in Section 4. Several examples are also discussed. Finally, Section 5 is devoted to present the main conclusions and to point out several issues that we are investigating.

\section{Simple materials}\label{sec:material}

\begin{definition}
A \textit{material body} is given by an oriented manifold $\mathcal{B}$ of dimension $3$ which may be embedded in $\mathbb{R}^{3}$. Points of $\mathcal{B}$ are called \textit{material points} or \textit{material particles} and will be denoted by capital letters ($X,Y,Z \in \mathcal{B}$)
\end{definition}
\noindent{Any open subset $\mathcal{U}$ of the manifold $\mathcal{B}$ is called a \textit{sub-body}.} A \textit{configuration} is an embedding $\phi : \mathcal{B} \rightarrow \mathbb{R}^{3}$. An \textit{infinitesimal configuration at a particle $X$} is given by the $1-$jet $j_{X,\phi \left(X\right)}^{1} \phi$ where $\phi$ is a configuration of $\mathcal{B}$. To study in detail the formalism of $1-$jets see \cite{SAUND}. A configuration, called \textit{reference configuration}, $\phi_{0}$ will be fixed. The open set $\mathcal{B}_{0} = \phi_{0} \left( \mathcal{B} \right)$ will be called \textit{reference state}. The local coordinates in the reference configuration will be denoted by $X^{I}$ and any other coordinates will be denoted by $x^{i}$.\\
A \textit{deformation} of the body $\mathcal{B}$ is defined as the change of configurations $\kappa = \phi_{1} \circ \phi_{0}^{-1}$ or, equivalently a diffeomorphism from the reference state $\mathcal{B}_{0}$ to any other open subset $\mathcal{B}_{1} = \phi_{1}\left( \mathcal{B} \right)$ of $\mathbb{R}^{3}$. Analogously, an \textit{infinitesimal deformation at $\phi_{0}\left(X\right)$} is given by a $1-$jet $j_{\phi_{0}\left(X\right) , \phi \left(X\right)}^{1} \kappa$ where $\kappa$ is a deformation.\\
Broadly speaking, the configuration is a way of manifesting the body into the ``\textit{real world}''. The points on the Euclidean space $\mathbb{R}^{3}$ will be called \textit{spatial points} and will be denoted by lower case letters ($x,y,z\in \mathbb{R}^{3}$).\\
Following the theory developed by W. Noll \cite{WNOLLTHE}, the internal properties of the body are characterized for the so-called \textit{constitutive equations}. For \textit{elastic simple bodies}, we will assume that the constitutive law depends on a material point only through the infinitesimal deformation at that point.
\begin{definition}
\rm
The \textit{mechanical response} of an (elastic) simple material $\mathcal{B}$, in a fixed reference configuration $\phi_{0}$, is formalized as a differentiable map $W$ from the set $\mathcal{B} \times Gl \left( 3 , \mathbb{R} \right)$, where $Gl \left( 3 , \mathbb{R} \right)$ is the general linear group of $3 \times 3$-regular matrices, to a fixed (finite dimensional) vector space $V$.
\end{definition}

In general, $V$ will be the space of \textit{stress tensors}. More particularly, the contact forces at a particle $X$ (in a fixed configuration $\phi$) are determined by a symmetric second-order tensor 
$$T_{X,\phi}: \mathbb{R}^{3} \rightarrow \mathbb{R}^{3}$$
on $\mathbb{R}^{3}$ called the \textit{stress tensor}. Then, the mechanical response is given as follows:
$$W \left( X , F \right) = T_{X,\phi},$$
where $F$ is the $1-$jet $j_{\phi_{0} \left( X \right),\phi\left( X\right)}^{1}\left( \phi \circ \phi_{0}^{-1}\right)$ at $\phi_{0} \left( X \right)$ of $\phi \circ \phi_{0}^{-1}$.\\
We should now introduce the \textit{rule of change of reference configuration}. In particular, consider  another configuration $\phi_{1}$ and the associated mechanical response $W_{1}$. Then, we will impose that,
\begin{equation}\label{1.5}
 W_{1} \left( X , F \right) = W \left( X , F \cdot C_{01} \right),
\end{equation}
for all regular matrix $F$, where $C_{01}$ is the associated matrix to the $1-$jet at $\phi_{0} \left( X \right)$ of $\phi_{1} \circ \phi_{0}^{-1}$. 
This fact is equivalent to the identity,
\begin{equation}\label{1.4}
 W \left( X , F_{0} \right) = W_{1} \left( X , F_{1} \right),
\end{equation}
for any configuration $\phi$, where $F_{i}$, $i=0,1$, is the associated matrix to the $1-$jet at $\phi_{i} \left( X \right)$ of $\phi \circ \phi_{i}^{-1}$ \cite{VMMDME}.\\

There are several equivalent ways of presenting the mechanical response. On the one hand, we may define define $W$ on the space of (local) configurations in such a way that for each configuration $\phi$ we have that
$$ W \left( j^{1}_{X, x} \phi \right) = W \left( X , F \right),$$
where $F$ is the associated matrix to the $1-$jet at $\phi_{0} \left( X \right)$ of $\phi \circ \phi_{0}^{-1}$. So, Eq. (\ref{1.4}) implies that this map does not depend on the chosen reference configuration.\\
On the other hand, consider $\Pi^{1}\left( \mathcal{B}, \mathcal{B}\right)$ the manifold of the $1-$jets of (local) diffeomorphisms from $\mathcal{B}$ to $\mathcal{B}$ (\cite{KMG}). Then, $W$ may be described as a differentiable map $W : \Pi^{1}\left( \mathcal{B} , \mathcal{B} \right) \rightarrow V$ from $\Pi^{1}\left( \mathcal{B} , \mathcal{B} \right)$ to the vector space $ V$ by the following identity,
\begin{equation}\label{4.1}
 W \left( j_{X,Y}^{1} \kappa\right) = W \left( X , F\right),
\end{equation}
where $F$ is the associated matrix to the $1-$jet at $\phi_{0} \left( X \right)$ of $\phi_{0}\circ \phi \circ \phi^{-1}_{0}$. If there is no danger of confusion, we will use in this paper these three ways of describing the mechanical response, indistinctly.\\
Observe that, restricting the mechanical response, any sub-body inherits the structure of elastic simple body from the body $\mathcal{B}$. A fundamental notion in the theory of W. Noll is the concept of \textit{material isomorphism}, which permits to compare the material properties of two different points.
\begin{definition}\label{1.33}
\rm
Let $\mathcal{B}$ be a body. Two material particles $X$ and $Y$ are materially isomorphic if, and only if, there exists a local diffeomorphism $\psi$ from an open subset $\mathcal{U} \subseteq \mathcal{B}$ of $X$ to an open subset $\mathcal{V} \subseteq \mathcal{B}$ of $Y$ such that $\psi \left(X\right) =Y$ and
\begin{equation}\label{4.2}
W \left( j^{1}_{Y, \kappa \left(Y\right)} \kappa \cdot j^{1}_{X,Y} \psi \right) = W \left( j^{1}_{Y, \kappa \left(Y\right)} \kappa\right),
\end{equation}
for all $j^{1}_{Y , \kappa \left(Y\right)} \kappa \in \Pi^{1}\left( \mathcal{B} , \mathcal{B} \right)$. Under these conditions, $j^{1}_{X,Y} \psi$ will be called a material \textit{isomorphism from $X$ to $Y$.} A material isomorphism from $X$ to itself is called a \textit{material symmetry}. In cases where it causes no confusion we often refer to the associated matrix $P$ as the material isomorphism (or symmetry).
\end{definition}

\begin{remark}
\rm
\noindent{We should notice that the elements of $\Pi^{1}\left( \mathcal{B} , \mathcal{B} \right)$ may be interpreted as linear isomorphisms $L_{X,Y}: T_{X}\mathcal{B}  \rightarrow T_{Y}\mathcal{B}$ beetwen the tangent spaces of the body $\mathcal{B}$ at two different particles $X$ and $Y$.}
\end{remark}
So, from a physical point of view, two points are materially isomorphic if their intrinsic properties are the same, i.e., they are part of the same material. In fact, we have that,
\begin{proposition}[\cite{VMMDME}]
Let $\mathcal{B}$ be a body. Two body points $X$ and $Y$ are materially isomorphic if, and only if, there exist two (local) configurations $\phi_{1}$ and $\phi_{2}$ such that
$$ W_{1} \left( X , F \right)= W_{2} \left( Y , F \right), \ \forall F,$$
where $W_{i}$ is the mechanical response associated to $\phi_{i}$ for $i=1,2$.
\end{proposition}

In the next sections we will interpret these concepts using a natural instrument, the \textit{material groupoid}. We will see how uniformity is intimately linked to the concept of transitivity of the groupoid.

\section{The structure of a double groupoid}\label{sec:structure}

In order to make the text as self-contained as possible, we will first introduce the general concept of groupoids, since double groupoids are, in some sense, special cases of groupoids although of greater internal complexity (see \cite{KMG} for a standard reference).\\

Let $ M$ be a set. A \textit{groupoid} over $M$ is a set $\Gamma$ with maps $\alpha,\beta : \Gamma \rightarrow M$ (\textit{source map} and \textit{target map} respectively), $\epsilon: M \rightarrow \Gamma$ (\textit{identities}), $i: \Gamma \rightarrow \Gamma$ (\textit{inversion map}) and $\cdot : \Gamma_{\left(2\right)} \rightarrow \Gamma$ (\textit{composition law})
where $\Gamma_{\left(k\right)} = \{ \left(g_{1}, \hdots , g_{k}\right) \in \Gamma \times \stackrel{k)}{\ldots} \times \Gamma  \; | \; \alpha\left(g_{i}\right)=\beta\left(g_{i+1}\right),  i=1, \hdots , k -1\}$, satisfying the following properties:
\begin{itemize}
\item[(1)] $\alpha$ and $\beta$ are surjective and for each $\left(g,h\right) \in \Gamma_{\left(2\right)}$,
$$ \alpha\left(g \cdot h \right)= \alpha\left(h\right), \ \ \ \beta\left(g \cdot h \right) = \beta\left(g\right).$$
\item[(2)] Associative law with the composition law, i.e.,
$$ g \cdot \left(h \cdot k\right) = \left(g \cdot h \right) \cdot k, \ \forall \left(g,h,k\right) \in \Gamma_{\left(3\right)}.$$
\item[(3)] For all $ g \in \Gamma$,
$$ g \cdot \epsilon \left( \alpha\left(g\right)\right) = g = \epsilon \left(\beta \left(g\right)\right)\cdot g .$$
Therefore,
$ \alpha \circ  \epsilon \circ \alpha = \alpha ,  \beta \circ \epsilon \circ \beta = \beta.$

\item[(4)] For each $g \in \Gamma$,
$$i\left(g\right) \cdot g = \epsilon \left(\alpha\left(g\right)\right) , \ \ \ g \cdot i\left(g\right) = \epsilon \left(\beta\left(g\right)\right).$$
Then,
$ \alpha \circ i = \beta ,  \beta \circ i = \alpha.$
\end{itemize}

Since $\alpha$ and $\beta$ are surjective, we have that
$$ \alpha \circ \epsilon = Id_{M}, \ \ \ \beta \circ \epsilon = Id_{M},$$
where the map $Id_{M}$ is the identity map at $M$.

The maps involved in the above definition are called \textit{structure maps}. The usual notation for a groupoid is $ \Gamma \rightrightarrows M$.

\medskip

$M$ is denoted by $\Gamma_{\left(0\right)}$ and it is identified with the set $\epsilon \left(M\right)$ of identities of $\Gamma$. $\Gamma$ is also denoted by $\Gamma_{\left(1\right)}$. The elements of $M$ are called \textit{objects} and the elements of $\Gamma$ are called \textit{morpishms}. Furthermore, for each $g \in \Gamma$ the element $i \left( g \right)$ is called \textit{inverse of $g$} and it is denoted by $g^{-1}$.

Let $\Gamma \rightrightarrows M$ be a groupoid. The map $\left(\alpha , \beta\right) : \Gamma \rightarrow M \times M$ is called the \textit{anchor map}. The space of sections of the anchor map is denoted by $\Gamma_{\left(\alpha, \beta\right)} \left(\Gamma\right)$.

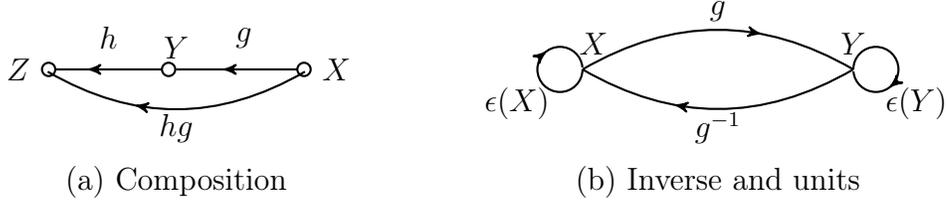
\begin{figure}[h!]
\begin{center}
\begin{tikzpicture} [scale=0.6]
\tikzset{->-/.style={decoration={
  markings,
  mark=at position .65 with {\arrow{stealth'}}},postaction={decorate}}}
  
  \draw[thick, o-o,->-] (3,0) -- (0,0); 
  \draw [thick,o-, ->-] (6,0) -- (3,0);
  \draw[thick,->-](5.85,-0.05) to [bend left] (0.15,-0.05);
  \node[left] at (0,0) {$Z$};
  \node [above] at (3,0) {$Y$};
  \node[right] at (6,0) {$X$};
  \node at (1.5,.7) {$h$};
    \node at (4.5,.7) {$g$};
      \node at (3,-1.3) {$hg$};
\node at (3,-2.5) {(a) Composition};
\begin{scope}[xshift=12cm]
  \draw[thick, ->-] (0,0) to [bend left] (6,0); 

  \draw[thick,->-](6,0) to [bend left] (0,0);
  \node[above] at (0.25,0.1) {$X$};

  \node[above] at (6,0.1) {$Y$};
  \node at (3,1.3) {$g$};

      \node at (3,-1.3) {$g^{-1}$};

      \draw[thick, ->-] (0,0) arc (360:5:0.5);
          \draw[thick, ->-] (6,0) arc (180:-175:0.5);
          \node[below] at (-1.45,-0.15) {$\epsilon(X)$};
            \node[below] at (7.4,-0.15) {$\epsilon(Y)$};
\node at (3,-2.5) {(b) Inverse and units};
\end{scope}
\end{tikzpicture}
\end{center}
\caption{Groupoid operations}
\label{fig:inverse}
\end{figure}

Roughly speaking, a groupoid may be thought as a set of ``\textit{arrows}'' ($g, h \in \Gamma$) joining points ($X , Y , Z \in M$) next to a composition law with similar rules to the composition of maps, see \cref{fig:inverse}.

Let $\Gamma \rightrightarrows M$ be a groupoid with $\alpha$ and $\beta$ the source map and target map, respectively. For each $X \in M$, the set
$$\Gamma^{X}_{X}= \beta^{-1}\left(X\right) \cap \alpha^{-1}\left(X\right),$$
is called the \textit{isotropy group of} $\Gamma$ at $X$. The set
$$\mathcal{O}\left(x\right) = \beta\left(\alpha^{-1}\left(X\right)\right) = \alpha\left(\beta^{-1}\left( X\right)\right),$$

is called the \textit{orbit} of $X$, or \textit{the orbit} of $\Gamma$ through $X$. The \textit{orbit space} of $\Gamma$ is the space of orbits of $\Gamma $ on $M$
Notice that the orbit of a point $X$ consists of the points which are ``\textit{connected}" with $X$ by a morphism in the groupoid while the isotropy group is given by the morphisms connecting $X$ with $X$. Thus, the isotropy groups inherits a \textit{bona fide} group structure.

\begin{definition}[Transitivity]

Let $\Gamma \rightrightarrows M$ be a groupoid. If $\mathcal{O}\left(X\right) = M$ for all $X \in M$, or equivalently $\left(\alpha,\beta\right) : \Gamma  \rightarrow M \times M$ is a surjective map, the groupoid $\Gamma \rightrightarrows M$ is called \textit{transitive}. 
\end{definition}

Roughly speaking, transitivity may be interpreted as the existence of an ``\textit{arrow}'' connecting any two points in $M$. In other words, \textit{there are no isolated sets of points}.

The property of transitivity will play a central role throughout this paper, as it constitutes a fundamental aspect of its application to the constitutive theory of materials. The mathematical framework developed here relies heavily on the nuanced interplay between transitivity and the structural properties of the materials, making it a key component in the analysis and formulation of material behavior (\cite{VMMDME}). 

\begin{definition}[Total intransitivity]

Let $\Gamma \rightrightarrows M$ be a groupoid. If $\mathcal{O}\left(X\right) =\{X\}$, or equivalently $\beta^{-1}\left(X\right) = \alpha^{-1}\left( X \right)=\Gamma_{X}^{X}$, then $X$ is called a \textit{fixed point}. When every $X \in M$ is a fixed point, the groupoid $\Gamma \rightrightarrows M$ is called \textit{totally intransitive}.
\end{definition}

Furthermore, a subset $N$ of $M$ is called \textit{invariant} \index{Invariant} if it is a union of orbits.
Finally, the sets,
$$ \alpha^{-1} \left(X \right) = \Gamma_{X}, \ \ \ \ \ \beta^{-1} \left(X \right) = \Gamma^{X},$$
are called $\alpha-$\textit{fibre at} $X$ and $\beta-$\textit{fibre at $X$}, respectively.

\begin{example}
[Groups]

A group is a groupoid over a point. Indeed, let $G$ be a group and $e$ the identity element of $G$. Then, $G \rightrightarrows \{e\}$ is a groupoid, where the operation law of the groupoid, $\cdot$, is the operation in $G$.

\end{example}

\begin{example}[Pair groupoid]\label{pair_groupoid}

For any set $A$, we shall consider the product space $ A \times A$. Then, the maps,
\begin{itemize}
\item[] $\alpha \left(a,b\right) = a , \ \ \beta \left(a,b\right)=b, \ \forall \left(a,b\right) \in  A \times A$
\item[] $\left(c,b\right)\cdot\left(a,c\right)= \left(a,b\right), \ \forall \left( c,b\right),\left(a,c\right) \in  A \times A$
\item[] $ \epsilon \left(a\right) = \left(a,a\right), \ \forall a \in A$
\item[] $ \left(a,b\right)^{-1}=\left(b,a\right), \ \forall \left(a,b\right) \in  A \times A$
\end{itemize}
endow $A \times A$ with a structure of groupoid over $A$, called the \textit{pair groupoid}.

\end{example}

\begin{example}[Frame groupoid and 1-jets groupoid]
\label{frame_groupoid}
    
Let us consider a vector bundle $A$ on a manifold $M$. For each $Z\in M$, denote by $A_{Z}$ the fibre of $A$ over $Z$. Then, $\Phi \left(A\right)$ is the set of linear isomorphisms $L_{X,Y}: A_{X} \rightarrow A_{Y}$, for $X,Y \in M$ and it may be endowed with the structure of groupoid with the following structure maps,
\begin{itemize}
\item[(i)] $\alpha\left(L_{X,Y}\right) = X$
\item[(ii)] $\beta\left(L_{X,Y}\right) = Y$
\item[(iii)] $L_{Y,Z} \cdot G_{X,Y} = L_{Y,Z} \circ G_{X,Y}, \ L_{Y,Z}: A_{Y} \rightarrow A_{Z}, \ G_{X,Y}: A_{X} \rightarrow A_{Y}$
\end{itemize}
This groupoid is called the \textit{frame groupoid on $A$}. A particular relevant case arises when we choose $A$ equal to the tangent bundle $TM$ of $M$. In this latter case, the groupoid will be called \textit{1-jets groupoid on} $M$ and denoted by $\Pi^{1} \left(M,M\right)$. Notice that any isomorphism $L_{X,Y}: T_{X}M \rightarrow T_{Y}M$ may be written as a $1-$jet $j_{X,Y}^{1} \psi$ of a local diffeomorphism $\psi$ from $M$ to $M$ such that $\psi \left( X \right) = Y$. Recall that the $1-$jet $j_{X,Y}^{1} \psi$ may by represented by that induced tangent map $T_{X}\psi: T_{X}M \rightarrow T_{Y}M$. 
\end{example}

\begin{definition}
\rm

If $\Gamma_{1} \rightrightarrows M_{1}$ and $\Gamma_{2} \rightrightarrows M_{2}$ are two groupoids then a \textit{morphism of groupoids} from $\Gamma_{1} \rightrightarrows M_{1}$ to $\Gamma_{2} \rightrightarrows M_{2}$ consists of two maps $\Phi : \Gamma_{1} \rightarrow \Gamma_{2}$ and $\phi : M_{1} \rightarrow M_{2}$ satisfying the commutative relations of the following diagrams, 
 \begin{center}
\includegraphics[width=13cm]{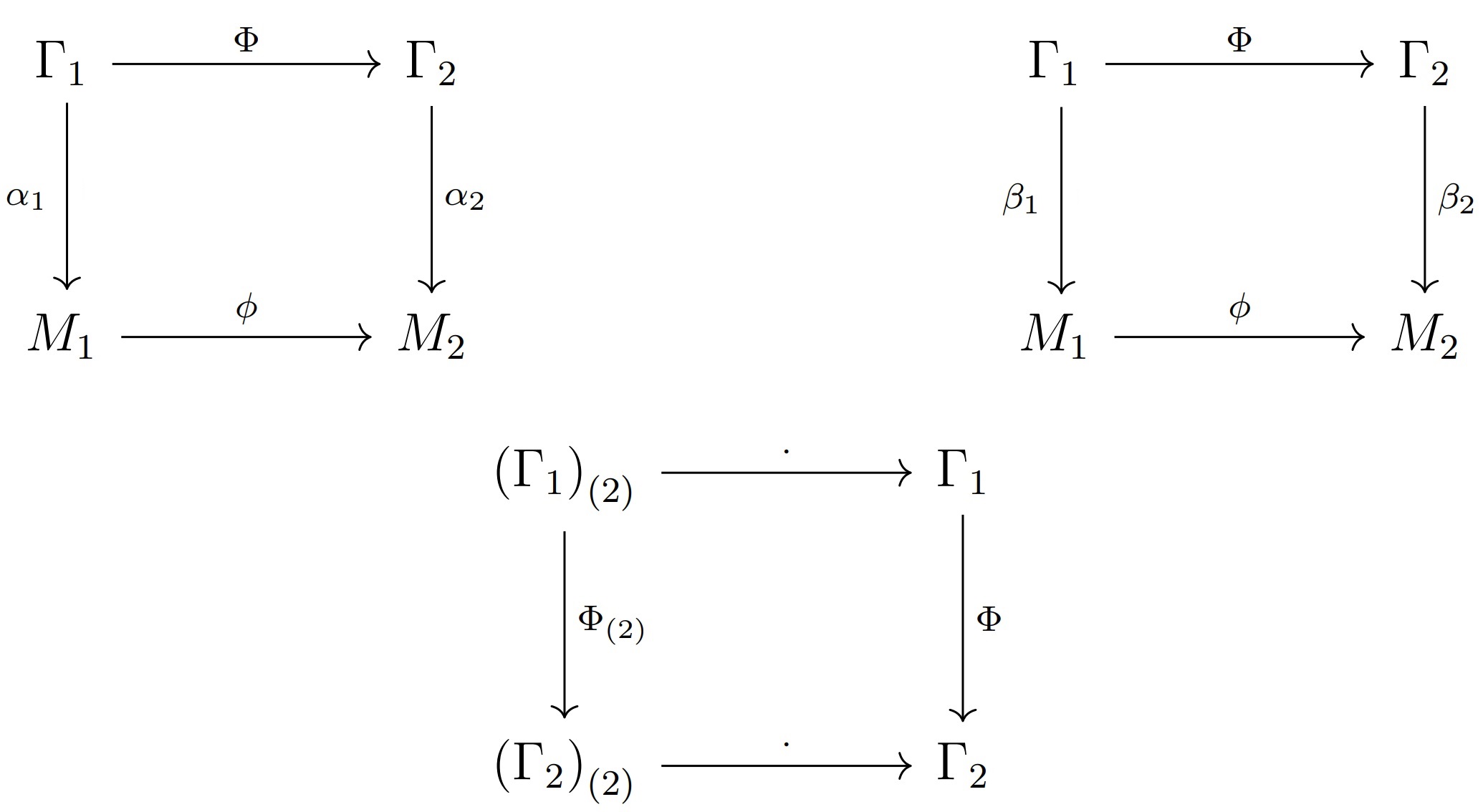}
\end{center}
where, 
$$\Phi_{(2)} \left( g_{1} , h_{1} \right) = \left( \Phi \left(g_{1} \right) ,  \Phi \left(h_{1} \right) \right)$$
for all $\left( g_{1} , h_{1} \right) \in \left(\Gamma_{1}\right)_{(2)}$. Equivalently, for any $g_{1} \in \Gamma_{1}$
\begin{equation}\label{4}
\alpha_{2} \left( \Phi \left(g_{1}\right)\right) = \phi \left(\alpha_{1} \left(g_{1} \right)\right), \ \ \ \ \ \ \ \beta_{2} \left( \Phi \left(g_{1}\right)\right) = \phi \left(\beta_{1} \left(g_{1} \right)\right),
\end{equation}
where $\alpha_{i}$ and $\beta_{i}$ are the source and the target maps of $\Gamma_{i} \rightrightarrows M_{i}$ respectively, for $i=1,2$, and preserves the composition, i.e.,
$$\Phi \left( g_{1} \cdot h_{1} \right) = \Phi \left(g_{1}\right) \cdot \Phi \left(h_{1}\right), \ \forall \left(g_{1} , h_{1} \right) \in \Gamma_{\left(2\right)}.$$

\end{definition}
Taking into account \cref{4}, $\phi$ is characterized by $\Phi$. Therefore, we will denote this morphism as $\Phi$, if there is no danger of confusion.

\begin{definition}[Subgroupoid]
Let $\Gamma \rightrightarrows M$ be a groupoid. Then, a \textit{subgroupoid} is another groupoid $\overline{\Gamma} \rightrightarrows \overline{M}$ such that $\overline{\Gamma} \subseteq \Gamma$, $\overline{M}\subseteq M$, and the inclusion maps $i_{\overline{\Gamma}}:\overline{\Gamma} : \hookrightarrow \Gamma$ and $i_{\overline{M}}:\overline{M} : \hookrightarrow M$ induce a morphism of groupoids.
\end{definition}

In other words, a subgroupoid of a given groupoid $\Gamma$ is a groupoid contained in $\Gamma$ with the ``\textit{same}'' structure maps (the restriction of the structure maps of $\Gamma$ to the correspondent sets of $\overline{\Gamma}$).

\vspace{0.5cm}

The definition of a double groupoid is somewhat more intricate. A {\it double groupoid} consists of a set $\mathcal Q$ endowed with two groupoid structures:
\begin{itemize}
    \item {\it Horizontal structure}: ${\mathcal Q} \rightrightarrows {\mathcal H}$ 
    
    \item {\it Vertical structure}: ${\mathcal Q} \rightrightarrows {\mathcal V}$ 
\end{itemize}

in such a way that each of the base spaces is itself a groupoid over a common base $\mathcal B$:

\begin{itemize}
    \item {\it Horizontal side groupoids}: ${\mathcal H}\rightrightarrows {\mathcal B}$

    \item {\it Vertical side groupoids}: ${\mathcal V}\rightrightarrows {\mathcal B}$
\end{itemize}

subject to a \textit{compatibility condition} between the horizontal and vertical structures, which ensures that the horizontal and vertical groupoid structure maps are morphisms with respect to each other. The precise meaning of this compatibility condition will be clarified later.

We will use a common notation for a double groupoid as the following diagram shows:
\begin{equation} \label{dg1}
\begin{matrix}
{\mathcal Q}                   &\rightrightarrows  & {\mathcal V} \\
  \downdownarrows    &                      &  \downdownarrows  \\
  {\mathcal H}                 & \rightrightarrows   &\mathcal{B}
\end{matrix}
\end{equation}
Given that the definition involves four distinct groupoids, it is essential to establish a clear notation for the elements, maps, and operations associated with each. Different authors approach this task in various ways, adopting notations that best suit their specific applications. The notations used here represent a combination of those found in \cite{ehresmann}, \cite{brown1}, and \cite{natale}. 

The elements of $\mathcal{B}$ will be denoted by capital letters $(X, Y, \dots)$. The arrows, units, and projections of the horizontal groupoid ${\mathcal{H}} \rightrightarrows {\mathcal{B}}$ will follow the standard notation for groupoids introduced earlier. For the vertical groupoid ${\mathcal{V}} \rightrightarrows {\mathcal{B}}$, the same notation will be employed, with the addition of a hat (circumflex accent) to distinguish it.

Products in ${\mathcal{H}} \rightrightarrows {\mathcal{B}}$ and ${\mathcal{V}} \rightrightarrows {\mathcal{B}}$ will be indicated by simple juxtaposition, as no confusion arises in this context. For example, $gh$ for arrows in $\mathcal{H}$, while ${\hat g}{\hat h}$ is used for arrows in $\mathcal{V}$. The source and target projections and units will similarly be distinguished by the presence or absence of a hat to avoid any ambiguity. Specifically, $\alpha(g)$ denotes the source of $g \in {\mathcal{H}}$, whereas ${\hat \alpha}({\hat g})$ denotes the source of ${\hat g} \in {\mathcal{V}}$.

The elements of $\mathcal{Q}$ will be denoted by double-strike letters $({\mathbb{G}}, {\mathbb{H}}, \dots)$ and will be graphically represented as squares, analogous to how the elements of $\mathcal{H}$ and $\mathcal{V}$ are represented as arrows. Each square consists of two horizontal parallel sides belonging to $\mathcal{H}$ (source and target maps of the horizontal structure, respectively) and two vertical sides belonging to $\mathcal{V}$ (source and target maps of the vertical structure, respectively). Furthermore, there are two distinct products defined for squares, corresponding to the two groupoid structures ${\mathcal{Q}} \rightrightarrows {\mathcal{V}}$ and ${\mathcal{Q}} \rightrightarrows {\mathcal{H}}$, which are denoted by $\hor$ and $\ver$, respectively.

\begin{equation} \label{dg2}
\begin{tikzpicture}[baseline=(current  bounding  box.center)]
  \draw[thick,-stealth'] (3,0) -- (0,0);
  \draw[thick,-stealth'] (3,3) -- (0,3);
  \draw[thick,-stealth'] (3,0) -- (3,3);
  \draw[thick,-stealth'] (0,0) -- (0,3);
\node at (1.5,1.5) {$\mathbb G$};
\node at (4.2,1.5) {${\hat g}={\tilde \alpha}_V({\mathbb G})$};
\node at (-1.3,1.5) {${\hat h}={\tilde \beta}_V({\mathbb G})$};
\node[above] at (1.5,3) {$h={\tilde \beta}_H({\mathbb G})$};
\node[below] at (1.5,0) {$g={\tilde \alpha}_H({\mathbb G})$};
\end{tikzpicture}
\end{equation}

The projections and units in each of the two groupoid structures of $\mathcal{Q}$ are being denoted with a superimposed tilde and a subscript $_H$ or $_V$, corresponding to the horizontal and vertical structures, respectively. Note that the codomain of the source and target maps of the horizontal structure is $\mathcal{V}$, while the codomain of the vertical structure is $\mathcal{H}$.

Let us now clarify the mentioned compatibility condition. On the one hand, the source and target maps os the horizontal structure are morphisms of groupoids and, therefore, as a first consecuence:
\begin{itemize}
    \item $\alpha \circ  \tilde{\alpha}_{\mathcal{H}} = \hat{\alpha} \circ \tilde{\alpha}_{\mathcal{V}} \ , \ \ \ \ \ \ \ \ \ \beta \circ  \tilde{\alpha}_{\mathcal{H}} = \hat{\alpha} \circ \tilde{\beta}_{\mathcal{V}}$

    \item $\alpha \circ \circ \tilde{\beta}_{\mathcal{H}} = \hat{\beta} \circ \tilde{\alpha}_{\mathcal{V}} \ , \ \ \ \ \ \ \ \ \ \beta \circ \circ \tilde{\beta}_{\mathcal{H}} = \hat{\beta} \circ \tilde{\beta}_{\mathcal{V}}$
\end{itemize}

These identities determine the coherence of the \textit{corners}'' and the prescribed directions of the \textit{arrows}'' in the diagram (\ref{dg2}). On the other hand, by using that the product of the horizontal structure is also a morphism of groupoids we have that, for each ${\mathbb G}, {\mathbb G}' \in \mathcal{Q}$, with $\tilde{\beta}_{\mathcal{V}}\left({\mathbb G}'\right) = \tilde{\alpha}_{\mathcal{V}} \left({\mathbb G} \right)$,

\begin{itemize}
    \item $\tilde{\alpha}_{\mathcal{H}}\left( {\mathbb G}\hor {\mathbb G}' \right) =  \tilde{\alpha}_{\mathcal{H}}\left( {\mathbb G}\right)  \tilde{\alpha}_{\mathcal{H}}\left({\mathbb G}' \right)$

    \item $\tilde{\beta}_{\mathcal{H}}\left( {\mathbb G}\hor {\mathbb G}' \right) =  \tilde{\beta}_{\mathcal{H}}\left( {\mathbb G}\right)  \tilde{\beta}_{\mathcal{H}}\left({\mathbb G}' \right)$
\end{itemize}

Graphically, these identities are represented as follows,
\begin{equation} \label{dg3}
\begin{tikzpicture}[baseline=(current  bounding  box.center)]
\foreach \x in {0,1.5,4.5}
{  \draw[thick,-stealth'] (1.5+\x,0) -- (0+\x,0);
  \draw[thick,-stealth'] (1.5+\x,1.5) -- (0+\x,1.5);
  \draw[thick,-stealth'] (1.5+\x,0) -- (1.5+\x,1.5);
  \draw[thick,-stealth'] (0+\x,0) -- (0+\x,1.5);}
\node[left] at(-0.,0.75) {${\mathbb G}\hor {\mathbb G}'\;=\;{\hat h}$};
\node[above] at(0.75,1.5) {$  h$};
\node[above] at(2.25,1.5) {$ h'$};
\node[below] at(0.75,-0.1) {$  g$};
\node[below] at(2.25,0) {$ g'$};
\node[right] at (3,0.75) {${\hat g}'\;=\;{\hat h}$};
\node[above] at (5.25,1.5) {$hh'$};
\node[below] at (5.25,0) {$gg'$};
\node[right] at (6,0.75) {${\hat g}'$};
\end{tikzpicture}
\end{equation}

The analogous statements in the vertical structure are,
\begin{itemize}
    \item $\tilde{\alpha}_{\mathcal{V}}\left( {\mathbb G}\ver {\mathbb G}' \right) =  \tilde{\alpha}_{\mathcal{V}}\left( {\mathbb G}\right)  \tilde{\alpha}_{\mathcal{V}}\left({\mathbb G}' \right)$

    \item $\tilde{\beta}_{\mathcal{V}}\left( {\mathbb G}\ver {\mathbb G}' \right) =  \tilde{\beta}_{\mathcal{V}}\left( {\mathbb G}\right)  \tilde{\beta}_{\mathcal{V}}\left({\mathbb G}' \right)$
\end{itemize}
Graphically,
\begin{equation} \label{dg4}
\begin{tikzpicture}[baseline=(current  bounding  box.center)]
\foreach \x in {0,-1.5}
{  \draw[thick,-stealth'] (1.5,0+\x) -- (0,0+\x);
  \draw[thick,-stealth'] (1.5,1.5+\x) -- (0,1.5+\x);
  \draw[thick,-stealth'] (1.5,0+\x) -- (1.5,1.5+\x);
  \draw[thick,-stealth'] (0,0+\x) -- (0,1.5+\x);}
\node[left] at(-0.,0) {${\mathbb G}\ver {\mathbb G}'\;=\;\;\;\;\;$};
\node[left] at (0,0.75) {${\hat h}$};
\node[above] at(0.75,1.5) {$  h$};
\node[right] at (1.5,0.75) {${\hat g}$};
\node[right] at (1.5,-0.75) {${\hat g}'$};
\node[below] at(0.75,-1.5) {$g'$};
\node[left] at (0,-0.75) {$\hat h'$};
\node[right] at (1.8,0.1) {$\;\;=\;\;\;{\hat h} {\hat h}'$};
\foreach \x in {3.5}
{\draw[thick,-stealth'] (1.5+\x,-0.75) -- (0+\x,0-0.75);
  \draw[thick,-stealth'] (1.5+\x,1.5-0.75) -- (0+\x,1.5-0.75);
  \draw[thick,-stealth'] (1.5+\x,0-0.75) -- (1.5+\x,1.5-0.75);
  \draw[thick,-stealth'] (0+\x,0-0.75) -- (0+\x,1.5-0.75);}
\node[right] at (5,0.1) {${\hat g}{\hat g}'$};
\node[above] at (4.25,0.75) {$h$};
\node[below] at (4.25,-0.75) {$g'$};
\end{tikzpicture}
\end{equation}

The unit square of the horizontal and vertical structures, respectively, look as follows:
\begin{equation} \label{dg3a}
\begin{tikzpicture}[baseline=(current  bounding  box.center)]
\foreach \x in {0}
{  \draw[thick] (-6+\x,0) -- (-7.5+\x,0);
\draw[thick] (-6+\x,-0.1) -- (-7.5+\x,-0.1);
  \draw[thick] (-6+\x,1.5) -- (-7.5+\x,1.5);
    \draw[thick] (-6+\x,1.6) -- (-7.5+\x,1.6);

  \draw[thick,-stealth'] (-6+\x,0) -- (-6+\x,1.5);
  \draw[thick,-stealth'] (-7.5+\x,0) -- (-7.5+\x,1.5);}
\node[left] at(-7.5,0.75) {${\tilde\epsilon}_V({\hat g})\;=\;{\hat g}$};
\node[above] at(-6.7,1.6) {$ \epsilon({\hat \beta}({\hat g}))$};
\node[below] at(-6.7,-0.2) {$ \epsilon({\hat \alpha}({\hat g}))$};
\node[left] at(-5.5,0.75) {${\hat g}$};
\node[right] at (1.5,0.75) {};

\foreach \x in {0}
{  \draw[thick,-stealth'] (0+\x,0) -- (-1.5+\x,0);
  \draw[thick,-stealth'] (0+\x,1.5) -- (-1.5+\x,1.5);
  \draw[thick] (0+\x,0) -- (0+\x,1.5);
  \draw[thick] (-1.5+\x,0) -- (-1.5+\x,1.5);
  \draw[thick] (0.1+\x,0) -- (0.1+\x,1.5);
  \draw[thick] (-1.6+\x,0) -- (-1.6+\x,1.5);}
\node[left] at(-1.5,0.75) {${ {\hat\epsilon}({ \beta}({ g}))}$};
\node[left] at(-3,0.75) {${\tilde\epsilon}_H({g})\;=$};
\node[above] at(-0.75,1.5) {$ g$};
\node[below] at(-0.75,-0.1) {$ g$};
\node[right] at (0.1,0.75) {${ {\hat\epsilon}({ \alpha}({ g}))}$};
\end{tikzpicture}
\end{equation}
for each $g \in \mathcal{H}$ and $\hat{g}\in \mathcal{V}$. The arrows representing the identities in the squares of a double groupoid will be drawn as two parallel straight lines. 

Finally, the inverse maps are also groupoid morphisms and, hence, we have that,

\begin{equation} \label{dg3a.inverse}
\begin{tikzpicture}[baseline=(current  bounding  box.center)]
\foreach \x in {0}
{  \draw[thick,-stealth'] (-6+\x,0) -- (-7.5+\x,0);
  \draw[thick,-stealth'] (-6+\x,1.5) -- (-7.5+\x,1.5);
  \draw[thick,-stealth'] (-6+\x,0) -- (-6+\x,1.5);
  \draw[thick,-stealth'] (-7.5+\x,0) -- (-7.5+\x,1.5);}
\node[left] at(-7.5,0.75) {${\hat h}$};
\node[above] at(-6.7,1.5) {$ h$};
\node[below] at(-6.7,-0.1) {$ g$};
\node[left] at(-5.5,0.75) {${\hat g}$};
\node[right] at (1.5,0.75) {};

\draw[->] (-5,0.75) .. controls (-4.5,2) and (-4,-2) .. (-2.5,0.75);
  \node at (-3.5, 1.25) {$\tilde{i}_{\mathcal{H}}\left( \mathbb{G}\right) $};

\foreach \x in {0}
{  \draw[thick,-stealth'] (0+\x,0) -- (-1.5+\x,0);
  \draw[thick,-stealth'] (0+\x,1.5) -- (-1.5+\x,1.5);
  \draw[thick,-stealth'] (0+\x,0) -- (0+\x,1.5);
  \draw[thick,-stealth'] (-1.5+\x,0) -- (-1.5+\x,1.5);}
\node[left] at(-1.5,0.75) {$\hat{h}^{-1}$};
\node[above] at(-0.75,1.5) {$ g$};
\node[below] at(-0.75,-0.1) {$ h$};
\node[right] at (0,0.75) {$\hat{g}^{-1}$};
\end{tikzpicture}
\end{equation}

\begin{equation} \label{dg3a.inverse2}
\begin{tikzpicture}[baseline=(current  bounding  box.center)]
\foreach \x in {0}
{  \draw[thick,-stealth'] (-6+\x,0) -- (-7.5+\x,0);
  \draw[thick,-stealth'] (-6+\x,1.5) -- (-7.5+\x,1.5);
  \draw[thick,-stealth'] (-6+\x,0) -- (-6+\x,1.5);
  \draw[thick,-stealth'] (-7.5+\x,0) -- (-7.5+\x,1.5);}
\node[left] at(-7.5,0.75) {${\hat h}$};
\node[above] at(-6.7,1.5) {$ h$};
\node[below] at(-6.7,-0.1) {$ g$};
\node[left] at(-5.5,0.75) {${\hat g}$};
\node[right] at (1.5,0.75) {};

\draw[->] (-5,0.75) .. controls (-4.5,2) and (-4,-2) .. (-2.5,0.75);
  \node at (-3.5, 1.25) {$\tilde{i}_{\mathcal{V}}\left( \mathbb{G}\right) $};

\foreach \x in {0}
{  \draw[thick,-stealth'] (0+\x,0) -- (-1.5+\x,0);
  \draw[thick,-stealth'] (0+\x,1.5) -- (-1.5+\x,1.5);
  \draw[thick,-stealth'] (0+\x,0) -- (0+\x,1.5);
  \draw[thick,-stealth'] (-1.5+\x,0) -- (-1.5+\x,1.5);}
\node[left] at(-1.5,0.75) {$\hat{g}$};
\node[above] at(-0.75,1.5) {$ h^{-1}$};
\node[below] at(-0.75,-0.1) {$ g^{-1}$};
\node[right] at (0,0.75) {$\hat{h}$};
\end{tikzpicture}
\end{equation}

Finally, the compatibility condition between the two structures imposes that,
\begin{equation} \label{dg5}
({\mathbb G}\hor{\mathbb H})\ver({\mathbb A}\hor{\mathbb B})=({\mathbb G}\ver{\mathbb A})\hor({\mathbb H}\ver{\mathbb B}),
\end{equation}
whenever the operations are possible. This condition implies that, in a square formed by four smaller squares with matching edges in contact, the result is the same whether one first composes horizontally and then vertically, or vice versa. In other words, the large square depicted below is well-defined.

\begin{equation} \label{dg6}
\begin{tikzpicture}[baseline=(current  bounding  box.center)]
\foreach \x in {0,1.5}
{\foreach \y in {0,1.5}
{  \draw[thick,-stealth'] (1.5+\x,0+\y) -- (0+\x,0+\y);
  \draw[thick,-stealth'] (1.5+\x,1.5+\y) -- (0+\x,1.5+\y);
  \draw[thick,-stealth'] (1.5+\x,0+\y) -- (1.5+\x,1.5+\y);
  \draw[thick,-stealth'] (0+\x,0+\y) -- (0+\x,1.5+\y);}}
\node at (0.75,0.75) {$\mathbb A$};
\node at (0.75,2.25) {$\mathbb G$};
\node at (2.25,0.75) {$\mathbb B$};
\node at (2.25,2.25) {$\mathbb H$};
\end{tikzpicture}
\end{equation}

In this way, one may observe how the representation of the elements of $\mathcal{Q}$ as ``\textit{oriented squares}'' is well aligned with the intuition behind the defining properties of a double groupoid.\\

One might quickly notice that, in this case, defining a \textit{transitivity condition} is not straightforward. The central difficulty lies in the non-existence of a uniquely defined anchor map. In general, we may define some kind of ``\textit{filling conditions}'', which, in some sense, describes the process of completing a square from a given partially defined one. As a first example, in \cite{brown1}, author study a \textit{filling condition} described by the surjectivity of the so-called \textit{double source map} $\left(\tilde{\alpha}_{\mathcal{V}},\tilde{\alpha}_{\mathcal{H}}\right): \mathcal{Q} \rightarrow \mathcal{V}\times_{\tilde{\alpha},\alpha}\mathcal{H}$, where, $\mathcal{V}\times_{\tilde{\alpha},\alpha}\mathcal{H} \coloneqq \left\{ \left(\tilde{g},g\right) \in \mathcal{V}\times \mathcal{H}\ : \ \tilde{\alpha}\left( \tilde{g}\right) = \alpha \left( g \right)\right\}$. In other words, we are completing squares from two given arrows, one vertical and one horizontal (figure \ref{dg7}).

\begin{equation}
\begin{tikzpicture}[baseline=(current  bounding  box.center)]\label{dg7}
\draw[thick,-stealth'] (0,0)--(-2,0);
\draw[thick,-stealth'] (0,0)--(0,2);
\draw[thick,-stealth'] (4,0)--(2,0);
\draw[thick,-stealth'] (4,0)--(4,2);
\draw[thick,-stealth'] (4,2)--(2,2);
\draw[thick,-stealth'] (2,0)--(2,2);
\node at (1,1) {$\Longrightarrow$};
\end{tikzpicture}
\end{equation}

\begin{example}[Coarse double groupoid]\label{Matdo1}
A useful example of a double groupoid is the \textit{coarse double groupoid}, $\square({\mathcal{H}}, {\mathcal{V}})$, generated by two groupoids, $\mathcal{H}$ and $\mathcal{V}$, with a common base set $\mathcal{B}$.
This double groupoid consists of all the <<\textit{consistent squares}>> that can be formed using the two groupoids as sides. In other words, the elements of $\square({\mathcal{H}}, {\mathcal{V}})$ are given by the quadruples $\left( \left( g,h \right) , \left(\hat{g} , \hat{h}\right) \right)$, where $g,h \in \mathcal{H}$ and $\hat{g} , \hat{h} \in \mathcal{V}$ such that,
\begin{eqnarray*}
\tilde{\alpha}_{\mathcal{H}}\left(g \right) &=& \tilde{\alpha}_{\mathcal{V}}\left(\hat{g} \right), \ \ \tilde{\alpha}_{\mathcal{H}}\left(h \right) = \tilde{\beta}_{\mathcal{V}}\left(\hat{g} \right)\\ \tilde{\beta}_{\mathcal{H}}\left(g \right) &=& \tilde{\alpha}_{\mathcal{V}}\left(\hat{h} \right), \ \ \tilde{\beta}_{\mathcal{H}}\left(h \right) = \tilde{\beta}_{\mathcal{V}}\left(\hat{h} \right)
\end{eqnarray*}
Then, the \textit{vertical product} is given by,
$$\left( \left( g,h \right) , \left(\hat{s} , \hat{h}\right) \right) \hor \left( \left( g',h' \right) , \left(\hat{g}' , \hat{s}\right) \right) = \left( \left( gg',hh' \right) , \left(\hat{g}' , \hat{h}\right) \right),$$
which may be graphically represented as in \cref{dg3}. The \textit{horizontal product} $\ver$ is defined in an analogous way.
\end{example}

Notice that any double groupoid $\mathcal{Q}$, with horizontal and vertical structures given by $\mathcal{H}$ and $\mathcal{V}$, respectively, can naturally be mapped inclusively into the coarse double groupoid generated by $\mathcal{H}$ and $\mathcal{V}$ by the map,
$$
\begin{array}{rccl}
\left(\tilde{\alpha}_{\mathcal{H}}, \tilde{\beta}_{\mathcal{H}}, \tilde{\alpha}_{\mathcal{V}}, \tilde{\beta}_{\mathcal{V}}\right) \colon & \mathcal{Q} & \longrightarrow & \square \left({\mathcal{H}}, {\mathcal{V}}\right)\\
& \mathbb{G} & \longmapsto & \left(\left(\tilde{\alpha}_{\mathcal{H}}\left(\mathbb{G}\right), \tilde{\beta}_{\mathcal{H}}\left(\mathbb{G}\right)\right), \left( \tilde{\alpha}_{\mathcal{V}}\left(\mathbb{G}\right), \tilde{\beta}_{\mathcal{V}}\left(\mathbb{G}\right)\right)\right)
\end{array}
$$

\begin{definition}[Core groupoid]
The {\it core} $\mathcal{K}$ of a double groupoid $\mathcal{Q}$ consists of the collection of all squares $\mathbb{G}$ such that
$$ \tilde{\alpha}_{\mathcal{H}}\left(\mathbb{G}\right) = \epsilon \left( X \right) , \ \ \ \tilde{\alpha}_{\mathcal{V}}\left(\mathbb{G}\right) = \hat{\epsilon} \left( X \right).$$
\end{definition}

The core may be endowed with a canonical groupoid structure by regarding its arrows as pairs of arrows $\mathbb{G} = (g,\hat{g})$ with common source $X$ and target point $Y$. So, we may define $\tilde{\alpha}_{\mathcal{K}}\left( \mathbb{G}\right) = X$ and $\tilde{\beta}_{\mathcal{K}}\left( \mathbb{G}\right) = Y$, as the source and the target map of $\mathcal{K}$, respectively.

Thus, the composition is given by,
\begin{eqnarray*}
    \left( \left( \epsilon \left( X \right),h \right) , \left(\hat{\epsilon} \left( X \right) , \hat{h}\right) \right) &\cdot_{\mathcal{K}}&  \left( \left( \epsilon \left( Z \right),h' \right) , \left(\hat{\epsilon} \left( Z \right) , \hat{h}'\right) \right)\\ 
    &=&  \left( \left( \epsilon \left( Z \right),hh' \right) , \left(\hat{\epsilon} \left( Z \right) , \hat{h} \hat{h}'\right) \right)
\end{eqnarray*}
In other words, for any two composable elements $\mathbb{K}_{1}, \mathbb{K}_{2} \in \mathcal{K}$ we have that,
$$\mathbb{K}_{1} \cdot_{\mathcal{K}} \mathbb{K}_{2} =\left( \mathbb{K}_{1} \hor \tilde{\epsilon}_{\mathcal{H}}\left( \tilde{\beta}_{\mathcal{H}}\left( \mathbb{K}_{2}\right)\right)\right) \ver \mathbb{K}_{2}$$

In many cases, the core groupoid serves to characterize the double groupoid \cite{brown1}, although the physical identity of the original side groupoids is not retained when transitioning to the core.

\begin{example}[Commutative squares]\label{dg12}
Let $\mathcal{Q}$ be a double groupoid characterized by the following diagram,
\begin{equation*}
\begin{matrix}
{\mathcal Q}                   &\rightrightarrows  & {\mathcal V} \\
  \downdownarrows    &                      &  \downdownarrows  \\
  {\mathcal H}                 & \rightrightarrows   &\mathcal{B}
\end{matrix}
\end{equation*}
in such a way that $\mathcal V$ and $\mathcal H$ are subgroupoids of a given groupoid $\Gamma$.

We will define the \textit{commutative squares} of $\mathcal{Q}$ as the elements $\mathbb{G} \in Q$ in such a way that,
\begin{equation}\label{dg11}
\tilde{\beta}_{\mathcal{V}}\left( \mathbb{G}\right) \cdot \tilde{\alpha}_{\mathcal{H}}\left( \mathbb{G}\right) = \tilde{\beta}_{\mathcal{H}}\left( \mathbb{G}\right) \cdot \tilde{\alpha}_{\mathcal{V}}\left( \mathbb{G}\right)
\end{equation}
Roughly speaking, the sides of the square commutes by the composition of the groupoid $\Gamma$. Notice that condition (\ref{dg11}) may be equivalently defined starting from any corner of the squares, which shows the consistency of the definition. More precisely, all these equalities are equivalent:
\begin{enumerate}[i)]
    \item $\tilde{\beta}_{\mathcal{V}}\left( \mathbb{G}\right) \cdot \tilde{\alpha}_{\mathcal{H}}\left( \mathbb{G}\right) = \tilde{\beta}_{\mathcal{H}}\left( \mathbb{G}\right) \cdot \tilde{\alpha}_{\mathcal{V}}\left( \mathbb{G}\right)$

    \item $ \tilde{\alpha}_{\mathcal{H}}\left( \mathbb{G}\right) \cdot \tilde{\alpha}_{\mathcal{V}}\left( \mathbb{G}\right) ^{-1}= \tilde{\beta}_{\mathcal{V}}\left( \mathbb{G}\right)^{-1}\cdot \tilde{\beta}_{\mathcal{H}}\left( \mathbb{G}\right) $

    \item $\tilde{\beta}_{\mathcal{H}}\left( \mathbb{G}\right)^{-1} \cdot\tilde{\beta}_{\mathcal{V}}\left( \mathbb{G}\right)  =  \tilde{\alpha}_{\mathcal{V}}\left( \mathbb{G}\right) \cdot \tilde{\alpha}_{\mathcal{H}}\left( \mathbb{G}\right)^{-1}$

    \item $\tilde{\alpha}_{\mathcal{V}}\left( \mathbb{G}\right)^{-1} \cdot \tilde{\beta}_{\mathcal{H}}\left( \mathbb{G}\right)^{-1}   =  \tilde{\alpha}_{\mathcal{H}}\left( \mathbb{G}\right)^{-1} \cdot \tilde{\beta}_{\mathcal{V}}\left( \mathbb{G}\right)^{-1} $
\end{enumerate}
Apart from the coherence of the definition of commutative squares, this result immediately implies that \textit{if }$\mathbb{G}$\textit{ is a commutative square, the inverse} $\mathbb{G}^{-1}$\textit{ is also a commutative square}.

On the other hand, recall that the identities ${\tilde\epsilon}_V({\hat g})$ and ${\tilde\epsilon}_H({ g})$, with $\hat{g}\in \mathcal{V}$ and $g\in \mathcal{H}$, have two opposite sides equal and the other two sides equal to the respective identities (see figure (\ref{dg3a})). Thus, it is easy to show that both, ${\tilde\epsilon}_V({\hat g})$ and ${\tilde\epsilon}_H({ g})$, are commutative squares.

Let us now consider two commutative squares $\mathbb{G}, \mathbb{H} \in \mathcal{Q}$ such that $\tilde{\alpha}_{\mathcal{H}}\left( \mathbb{G}\right) = \tilde{\beta}_{\mathcal{H}}\left( \mathbb{H}\right)$. Then,
\begin{align*}
\tilde{\alpha}_{\mathcal{H}}\left(\mathbb{G}\ver \mathbb{H}\right) &=  \tilde{\alpha}_{\mathcal{H}}\left( \mathbb{H}\right)\\
\tilde{\beta}_{\mathcal{H}}\left(\mathbb{G}\ver \mathbb{H}\right) &=  \tilde{\beta}_{\mathcal{H}}\left( \mathbb{G}\right)\\
\tilde{\alpha}_{\mathcal{V}}\left(\mathbb{G}\ver \mathbb{H}\right) &=  \tilde{\alpha}_{\mathcal{V}}\left(\mathbb{G}\right) \cdot \tilde{\alpha}_{\mathcal{V}}\left( \mathbb{H}\right)\\
\tilde{\beta}_{\mathcal{V}}\left(\mathbb{G}\ver \mathbb{H}\right) &=  \tilde{\beta}_{\mathcal{V}}\left( \mathbb{G}\right) \cdot \tilde{\beta}_{\mathcal{V}}\left( \mathbb{H}\right)
\end{align*}
Thus,
\begin{align*}
   \tilde{\beta}_{\mathcal{V}}\left(\mathbb{G}\ver \mathbb{H}\right)\cdot  \tilde{\alpha}_{\mathcal{H}}\left(\mathbb{G}\ver \mathbb{H}\right) &=   \tilde{\beta}_{\mathcal{V}}\left( \mathbb{G}\right) \cdot \tilde{\beta}_{\mathcal{V}}\left( \mathbb{H}\right) \cdot \tilde{\alpha}_{\mathcal{H}}\left( \mathbb{H}\right)\\
   &= \tilde{\beta}_{\mathcal{V}}\left( \mathbb{G}\right) \cdot \tilde{\beta}_{\mathcal{H}}\left( \mathbb{H}\right) \cdot \tilde{\alpha}_{\mathcal{V}}\left( \mathbb{H}\right)  \\
 &= \tilde{\beta}_{\mathcal{V}}\left( \mathbb{G}\right) \cdot \tilde{\alpha}_{\mathcal{H}}\left( \mathbb{G}\right) \cdot \tilde{\alpha}_{\mathcal{V}}\left( \mathbb{H}\right)  \\
 &= \tilde{\beta}_{\mathcal{H}}\left( \mathbb{G}\right) \cdot \tilde{\alpha}_{\mathcal{V}}\left( \mathbb{G}\right) \cdot \tilde{\alpha}_{\mathcal{V}}\left( \mathbb{H}\right)  \\
&=   \tilde{\beta}_{\mathcal{H}}\left(\mathbb{G}\ver \mathbb{H}\right)\cdot \tilde{\alpha}_{\mathcal{V}}\left(\mathbb{G}\ver \mathbb{H}\right)
\end{align*}
i.e., $\mathbb{G}\ver \mathbb{H}$ is, again, a commutative square. Analogously, we prove that for any two commutative squares  $\mathbb{G}', \mathbb{H}' \in \mathcal{Q}$ such that $\tilde{\alpha}_{\mathcal{V}}\left( \mathbb{G}'\right) = \tilde{\beta}_{\mathcal{HV}}\left( \mathbb{H}'\right)$, it satisfies that $\mathbb{G}\hor \mathbb{H}$ is a commutative square. Therefore, let us denote by $\boxdot \mathcal{Q}$ the set of commutative squares of $\mathcal{Q}$. Then, the set $\boxdot \mathcal{Q}$ is a double groupoid characterized by the following diagram,
\begin{equation} \label{dg10}
\begin{matrix}
{\boxdot\mathcal Q}                   &\rightrightarrows  & {\mathcal V} \\
  \downdownarrows    &                      &  \downdownarrows  \\
  {\mathcal H}                 & \rightrightarrows   &\mathcal{B}
\end{matrix}
\end{equation}
which will be called the \textit{groupoid of commutative squares}.
\end{example}

\section{Material double groupoid}\label{sec:doublegrop}

Let us consider a simple material $\mathcal{B}$. Then, we may consider $\Pi^{1}\left( \mathcal{B}, \mathcal{B}\right)\rightrightarrows \mathcal{B}$ the groupoid of $1-$jets of local diffeomorphisms over the body manifold $\mathcal{B}$. In other words, the \textit{morphisms} of the groupoids will be given by $1-$jets $j_{X,Y}^{1}\phi \in \Pi^{1}\left( \mathcal{B}, \mathcal{B}\right)$, with $\phi$ a local diffemorphism from $\mathcal{B}$ to $\mathcal{B}$ such that $\phi \left( X \right) = Y$ (see example \ref{frame_groupoid}).\\
For any two material particles $X,Y  \in \mathcal{B}$, the collection of all material isomorphisms from $X$ to $Y$ will be denoted by $\Omega \left( \mathcal{B}\right)_{X}^{Y}$. Notice that the composition and inverse of material isomorphisms are material isomorphisms and, therefore, the set
\begin{equation}\label{materialgroupoid232}
\Omega \left( \mathcal{B}\right) = \cup_{X,Y \in \mathcal{B}} \Omega \left( \mathcal{B}\right)_{X}^{Y}. 
\end{equation}
is a subgroupoid of $\Pi^{1}\left( \mathcal{B} , \mathcal{B} \right)$. This groupoid will be called the \textit{material groupoid of} $\mathcal{B}$ (see \cite{VMMDME}).\\
The material symmetry group $\Omega \left( \mathcal{B}\right)_{X}^{X}$ at a body point $X \in \mathcal{B}$ is simply the isotropy group of $\Omega \left( \mathcal{B}\right)$ at $X$. For any $X\in \mathcal{B}$, the set of material isomorphisms from $X$ to any other point (resp. from any point to $X$) will be denoted by $\Omega\left( \mathcal{B}\right)_{X}$ (resp. $\Omega\left( \mathcal{B}\right)^{X}$). Finally, the structure maps of $\Omega \left( \mathcal{B} \right)$ will be denoted by $\alpha$, $\beta$, $\epsilon$ and $i$, because they are just the restrictions of the corresponding ones on $\Pi^{1} \left(  \mathcal{B} , \mathcal{B} \right)$.

\begin{proposition}\label{Matdo3}
Let $\mathcal{B}$ be a body. $\mathcal{B}$ is uniform if and only if $\Omega \left( \mathcal{B}\right)$ is a transitive subgroupoid of $\Pi^{1} \left( \mathcal{B} , \mathcal{B}\right)$.
\end{proposition}

Thus, uniformity is characterized by the transitivity of the material groupoid. In other words, \textit{uniformity is characterized by the surjectivity of the anchor map} $\left(\alpha , \beta \right) : \Omega \left(\mathcal{B}\right) \rightarrow \mathcal{B} \times \mathcal{B}$, where $\alpha$ and $\beta$ are the source and the target maps, respectively.\\\\

Consider now two uniform materials given by their transitive material groupoids $\Omega_{1}\left( \mathcal{B}\right)$ and $ \Omega_{2}\left( \mathcal{B}\right)$ over the body manifold $\mathcal{B}$. Then, we may define the coarse double groupoid generated by $\Omega_{1}\left( \mathcal{B}\right)$ and $ \Omega_{2}\left( \mathcal{B}\right)$ (see example \cref{Matdo1}) denoted by $\Box \left( \Omega_{1}\left( \mathcal{B}\right), \Omega_{2}\left( \mathcal{B}\right)\right)$, i.e., the double groupoid which consists of all the ``\textit{squares}'', that can be formed using the
arrows of $\Omega_{1}\left( \mathcal{B}\right)$ and $\Omega_{2}\left( \mathcal{B}\right)$ as sides.\\

Formally, the elements of $\Box \left( \Omega_{1}\left( \mathcal{B}\right), \Omega_{2}\left( \mathcal{B}\right)\right)$ are denoted by $\left(\left(g_{1},h_{1}\right), \left(g_{2},h_{2}\right)\right)$ such that,
\begin{equation}
\alpha \left( g_{1} \right) = \alpha \left( g_{2} \right), \ \ \alpha \left( h_{1} \right) = \beta \left( g_{2} \right), \ \ \beta \left( g_{1} \right) = \alpha \left( h_{2} \right), \ \ \beta \left( h_{1} \right) = \beta \left( h_{2} \right)
\end{equation}

Recall that the source and the target maps of $ \Omega_{1}\left( \mathcal{B}\right), \Omega_{2}\left( \mathcal{B}\right)$ are denoted by the same symbols, because they are both subgroupoids of the same groupoid $\Pi^{1}\left( \mathcal{B}, \mathcal{B}\right)$. Altough the expressions of the elements of $\Box \left( \Omega_{1}\left( \mathcal{B}\right), \Omega_{2}\left( \mathcal{B}\right)\right)$ is clear, to enhance the graphical intuition of the elements of the double material groupoid it is still useful to depicted as \textit{squares of arrows} as follows,
\begin{equation} \label{dg8}
\begin{tikzpicture}[baseline=(current  bounding  box.center)]
  \draw[thick,-stealth'] (3,0) -- (0,0);
  \draw[thick,-stealth'] (3,3) -- (0,3);
  \draw[thick,red,-stealth'] (3,0) -- (3,3);
  \draw[thick,red,-stealth'] (0,0) -- (0,3);
\node at (3.5,1.5) {${g_{2}}$};
\node at (-0.5,1.5) {${h_{2}}$};
\node[above] at (1.5,3) {$h_{1}$};
\node[below] at (1.5,0) {$g_{1}$};
\end{tikzpicture}
\end{equation}
where the \textit{vertical} will be red, representing the material isomorphisms of the second material ($\Omega_{2}\left( \mathcal{B}\right)$), meantime the \textit{horizontal} will be black, representing the material isomorphisms of the first material ($\Omega_{1}\left( \mathcal{B}\right)$).\\

A \textit{composite material} is modelized as a permanent combination of two or more (possibly) distinct materials. Although each component remains identifiable within the mixture, the composite itself behaves as a single body $\mathcal{B}$. In particular, the deformation and its gradient are identical for all components.

For the sake of simplicity, we will deal with the case of a \textit{binary composite}, i.e. a composite material given by the combination of two elastic materials, 
$\Omega_{1}\left( \mathcal{B}\right)$ and 
$\Omega_{2}\left( \mathcal{B}\right)$, each of which, when considered independently, constitutes a\textit{ uniform body}. Consequently, the associated material groupoids, are transitive subgroupoids or $\Pi^{1}\left( \mathcal{B}, \mathcal{B}\right)$ (\cref{Matdo3}).

Following \cite{eps01}, and assuming that the bonding of the materials does not involve any chemical reactions, it is reasonable to posit that the mechanical response of the composite results from a weighted combination of the responses of its constituent materials. Consequently, the material groupoid of the composite is given by the intersection $\Omega_{1}\left(\mathcal{B}\right) \cap \Omega_{2}\left(\mathcal{B}\right)$ of the material groupoids of the individual components. In more intuitive terms, the material groupoid of the composite consists of all arrows that are common to the material groupoids of its components, which is, again, a new groupoid.

Regarding the remaining common arrows, several scenarios are possible. At one extreme, it may be the case that, apart from the isotropy groups, no common arrows exist. In this situation, despite the transitivity of the component groupoids $\Omega_{1}\left(\mathcal{B}\right)$ and $\Omega_{2}\left(\mathcal{B}\right)$, the material groupoid of the composite is completely intransitive.

At the other extreme, if there exists at least one common arrow between every pair of points and, in this case, the material groupoid of the composite is transitive. From a physical perspective, the transitive case corresponds to a scenario in which the two components have been consistently blended in the same manner at all points of the body. Thus, the we have the following natural definition:

\begin{definition}[Uniformity of a composite]
A binary composite given by the materials $\Omega_{1}\left(\mathcal{B}\right)$ and $\Omega_{2}\left(\mathcal{B}\right)$ is uniform if the intersection of the material groupoids is a transitive groupoid.
\end{definition}

Notice that, when the composite is uniform, for any two material particles $X$ and $Y$, there exists an \textit{arrow} ($1-$jet of a material isomorphism from $X$ to $Y$) which is a material isomorphism for both materials. In terms of the material double groupoids, from two given identities at the same point $X$, one vertical and one horizontal, we may \textit{complete} the square, i.e.,

\begin{equation*} \label{dg9}
\begin{tikzpicture}[baseline=(current  bounding  box.center)]
\draw[thick] (0.5,0) -- (-2.5,0);
  \draw[thick] (0.5,-0.1) -- (-2.5,-0.1);
  \draw[thick,red] (0.5,0) -- (0.5,3);
  \draw[thick,red] (0.6,0) -- (0.6,3);
  \node at (1.3,1.5) {${\epsilon\left( X \right)}$};
\node[below] at (-1,-0.1) {$\epsilon\left( X \right)$};

\node at (3,1) {$\Longrightarrow$};

  \draw[thick] (8,0) -- (5,0);
  \draw[thick] (8,-0.1) -- (5,-0.1);
  \draw[thick,-stealth'] (8,3) -- (5,3);
  \draw[thick,red] (8,0) -- (8,3);
  \draw[thick,red] (8.1,0) -- (8.1,3);
  \draw[thick,,red,-stealth'] (5,0) -- (5,3);
\node at (8.7,1.5) {${\epsilon\left( X \right)}$};
\node at (4.7,1.5) {${h_{2}}$};
\node[above] at (6.5,3) {$h_{1}$};
\node[below] at (6.5,-0.1) {$\epsilon\left( X \right)$};
\end{tikzpicture}
\end{equation*}

which is commutative ($h_{2} \cdot \epsilon \left( \alpha \left(h_{2}\right)\right) = h_{1}\cdot \epsilon \left( \alpha \left( h_{1}\right)\right)$). So, we will only be interested on double groupoid of commutative squares (\cref{dg12}), in this case denoted by $\boxdot
 \left( \Omega_{1}\left( \mathcal{B}\right), \Omega_{2}\left( \mathcal{B}\right)\right)$. Therefore, an element $\left(\left(g_{1},h_{1}\right), \left(g_{2},h_{2}\right)\right) \in \Box \left( \Omega_{1}\left( \mathcal{B}\right), \Omega_{2}\left( \mathcal{B}\right)\right)$ is in $\boxdot \left( \Omega_{1}\left( \mathcal{B}\right), \Omega_{2}\left( \mathcal{B}\right)\right)$ if, and only if,
\begin{equation}\label{Matdo2}
h_{2} \cdot g_{1} \ = \ h_{1} \cdot g_{2}
\end{equation}
Roughly speaking, both ``\textit{paths}'' from $\alpha \left( g_{1}\right) = \alpha \left( g_{2} \right)$ to $\beta\left(h_{2}\right) = \beta\left(h_{1}\right)$ coincide.

\begin{definition}[Composite groupoid]
Let be two uniform materials given by their transitive material groupoids $\Omega_{1}\left( \mathcal{B}\right), \Omega_{2}\left( \mathcal{B}\right)$ over the body manifold $\mathcal{B}$. The \textbf{\textit{material groupoid of the composite}} will be given by the double groupoid $\boxdot \left( \Omega_{1}\left( \mathcal{B}\right), \Omega_{2}\left( \mathcal{B}\right)\right)$ of commuting squares.
\end{definition}

It is important to highlight that the material groupoid of the composite \textit{encapsulates} the elastic properties of both materials, and those of the composite, in a singular algebraic structure. In this case, we will denote the structural maps of horizontal structure (resp. vertical structure) by $\tilde{\alpha}_{1}, \ \tilde{\beta}_{1}, \ \dots$ (resp. $\tilde{\alpha}_{2}, \ \tilde{\beta}_{2}, \ \dots$).

\begin{proposition}
Let $\boxdot \left( \Omega_{1}\left( \mathcal{B}\right), \Omega_{2}\left( \mathcal{B}\right)\right)$ the material groupoid of the composite. Then, the composite is uniform if, and only if, the map  $\left(\beta \circ\tilde{\beta}_{2},\tilde{\alpha}_{2},\tilde{\alpha}_{1}\right): \boxdot \left( \Omega_{1}\left( \mathcal{B}\right), \Omega_{2}\left( \mathcal{B}\right)\right)   \rightarrow \mathcal{B}\times \Omega_{2}\left( \mathcal{B}\right)\times_{\tilde{\alpha},\alpha}\Omega_{1}\left( \mathcal{B}\right)$, where, $\Omega_{2}\left( \mathcal{B}\right)\times_{\tilde{\alpha},\alpha}\Omega_{1}\left( \mathcal{B}\right) \coloneqq \left\{ \left(\tilde{g},g\right) \in\Omega_{2}\left( \mathcal{B}\right)\times \Omega_{1}\left( \mathcal{B}\right)\ : \ \alpha\left( \tilde{g}\right) = \alpha \left( g \right)\right\}$, satisfies that,
\begin{equation}\label{dg13}
\mathcal{B}\times \epsilon\left( \mathcal{B}\right) \times \epsilon\left( \mathcal{B}\right) \subseteq\left(\beta \circ\tilde{\beta}_{2},\tilde{\alpha}_{2},\tilde{\alpha}_{1}\right) \left(\boxdot \left( \Omega_{1}\left( \mathcal{B}\right), \Omega_{2}\left( \mathcal{B}\right)\right) \right)
\end{equation}

\end{proposition}

Thus, the condition of being uniform is not exactly given by the surjectivity of the map $\beta \circ\tilde{\beta}_{2},\tilde{\alpha}_{2},\tilde{\alpha}_{1}$, but a weaker condition \cref{dg13}.

Recall that, for the case of one simple material, uniformity is characterized by the transitivity of the material groupoid (\cref{Matdo3}). In the case of the material groupoid of the composite, we have a structure of double groupoid and, as a consequence, we have available more than one possible notion of ``\textit{transitivity}''.

In this paper, we will use this \textit{degree of freedom} to study the uniformity of the composite, as well as, to present new notions of \textit{compatibility} between the elastic properties of the materials, relating them with the mentioned uniformity of the composite.

\vspace{1cm}

\subsection*{Horizontal and vertical transitivity}

Two of the most common ``\textit{transitivity properties}'' in the theory of double groupoids are the so-called \textit{horizontal} and \textit{vertical transitivity}.

\begin{definition}[Horizontal property]
Let be two uniform materials given by their transitive material groupoids $\Omega_{1}\left( \mathcal{B}\right)$ and $ \Omega_{2}\left( \mathcal{B}\right)$ over the body manifold $\mathcal{B}$. We will say that the material groupoid of the composite $\boxdot \left( \Omega_{1}\left( \mathcal{B}\right), \Omega_{2}\left( \mathcal{B}\right)\right)$ is \textit{\textbf{horizontally transitive}} if the map $\left(\tilde{\alpha}_{2}, \tilde{\alpha}_{1}, \tilde{\beta}_{2}\right): \boxdot \left( \Omega_{1}\left( \mathcal{B}\right), \Omega_{2}\left( \mathcal{B}\right)\right) \to
\Omega_{2}\left( \mathcal{B}\right)\times_{ \alpha, \alpha } \Omega_{1}\left( \mathcal{B}\right)\times_{\beta, \alpha} \Omega_{2}\left( \mathcal{B}\right) $ is surjective, where $ \Omega_{2}\left( \mathcal{B}\right)\times_{ \alpha, \alpha } \Omega_{1}\left( \mathcal{B}\right)\times_{\beta, \alpha} \Omega_{2}\left( \mathcal{B}\right) $ is given by the triples $\left(g_{2},g_{1},h_{2}\right) \in \Omega_{2}\left( \mathcal{B}\right) \times \Omega_{1}\left( \mathcal{B}\right)\times \Omega_{2}\left( \mathcal{B}\right)$ such that $\alpha \left( g_{1}\right) = \alpha \left( g_{2}\right)$ and $\beta \left( g_{1}\right) = \alpha \left( h_{2}\right)$.
\end{definition}

\begin{definition}[Vertical transitivity]

Let be two uniform materials given by their transitive material groupoids $\Omega_{1}\left( \mathcal{B}\right)$ and $ \Omega_{2}\left( \mathcal{B}\right)$ over the body manifold $\mathcal{B}$. We will say that the material groupoid of the composite $\boxdot \left( \Omega_{1}\left( \mathcal{B}\right), \Omega_{2}\left( \mathcal{B}\right)\right)$ is \textit{\textbf{vertically transitive}} if the map $\left(\tilde{\alpha}_{1} , \tilde{\alpha}_{2}, \tilde{\beta}_{1}\right): \boxdot \left( \Omega_{1}\left( \mathcal{B}\right), \Omega_{2}\left( \mathcal{B}\right)\right) \to \Omega_{1}\left( \mathcal{B}\right)\times_{\alpha, \alpha} \Omega_{2}\left( \mathcal{B}\right)\times_{\beta, \alpha} \Omega_{1}\left( \mathcal{B}\right) $ is surjective, where $\Omega_{1}\left( \mathcal{B}\right)\times_{\alpha, \alpha} \Omega_{2}\left( \mathcal{B}\right)\times_{\beta, \alpha} \Omega_{1}\left( \mathcal{B}\right)$ is given by the triples $\left(g_{1},g_{2},h_{1}\right) \in \Omega_{1}\left( \mathcal{B}\right) \times \Omega_{2}\left( \mathcal{B}\right)\times \Omega_{1}\left( \mathcal{B}\right)$ such that $\alpha \left( g_{2}\right) = \alpha \left( g_{1}\right)$ and $\beta \left( g_{2}\right) = \alpha \left( h_{1}\right)$.
\end{definition}

Pictorially, the horizontal transitivity (resp. vertical transitive) is equivalent to the imposition of the existence of a box with any given three sides. If the missing side is horizontal (vertical) the transitivity is horizontal (vertical) (see picture below). \\

\begin{equation*} \label{dg14}
\begin{tikzpicture}[baseline=(current  bounding  box.center)]
\draw[thick,-stealth'] (0.5,0) -- (-2.5,0);
  \draw[thick,red,-stealth'] (0.5,0) -- (0.5,3);
  \draw[thick,red,-stealth'] (-2.5,0) -- (-2.5,3);
  \node at (-3,1.5) {${h_{2}}$};
  \node at (1,1.5) {${g_{2}}$};
\node[below] at (-1,-0.1) {$g_{1}$};

\node at (3,1.5) {\small{Hor. transit.}};

\node at (3,1) {$\Longrightarrow$};

  \draw[thick] (8,0) -- (5,0);
  
  \draw[thick,-stealth'] (8,3) -- (5,3);
  \draw[thick,red,-stealth'] (8,0) -- (8,3);
  \draw[thick,,red,-stealth'] (5,0) -- (5,3);
\node at (8.4,1.5) {${g_{2}}$};
\node at (4.6,1.5) {${h_{2}}$};
\node[above] at (6.5,3) {$h_{1}$};
\node[below] at (6.5,-0.1) {$g_{1}$};
\end{tikzpicture}
\end{equation*}

\begin{equation*} \label{dg15}
\begin{tikzpicture}[baseline=(current  bounding  box.center)]
\draw[thick,-stealth'] (0.5,0) -- (-2.5,0);
  \draw[thick,red,-stealth'] (0.5,0) -- (0.5,3);
  \draw[thick,-stealth'] (0.5,3) -- (-2.5,3);
\node[above] at (-1,3) {$h_{1}$};
\node at (1,1.5) {${g_{2}}$};
\node[below] at (-1,-0.1) {$g_{1}$};

\node at (3,1.5) {\small{Ver. transit.}};

\node at (3,1) {$\Longrightarrow$};

  \draw[thick,-stealth'] (8,0) -- (5,0);
  
  \draw[thick,-stealth'] (8,3) -- (5,3);
  \draw[thick,red,-stealth'] (8,0) -- (8,3);
  \draw[thick,,red,-stealth'] (5,0) -- (5,3);
\node at (8.4,1.5) {${g_{2}}$};
\node at (4.6,1.5) {${h_{2}}$};
\node[above] at (6.5,3) {$h_{1}$};
\node[below] at (6.5,-0.1) {$g_{1}$};
\end{tikzpicture}
\end{equation*}

Notice that, the condition of horizontal transitivity may be equivalently expressed as follows: for any triple $\left(g_{2},g_{1},h_{2}\right) \in \Omega_{2}\left( \mathcal{B}\right)\times_{ \alpha, \alpha } \Omega_{1}\left( \mathcal{B}\right)\times_{\beta, \alpha} \Omega_{2}\left( \mathcal{B}\right)$, it satisfies that
\begin{equation}\label{Matdo4}
 h_{2}\cdot g_{1} \cdot   g_{2}^{-1} \in \Omega_{1}\left(\mathcal{B}\right)
\end{equation}
So, we may easily prove the following result
\begin{proposition}
Let be two uniform materials given by their transitive material groupoids $\Omega_{1}\left( \mathcal{B}\right)$ and $ \Omega_{2}\left( \mathcal{B}\right)$ over the body manifold $\mathcal{B}$. Then, the material groupoid of the composite is horizontally transitive if, and only if,
$$ h_{2}\cdot \Omega_{1}\left(\mathcal{B}\right)_{X}^{Y} \cdot   g_{2}  = \Omega_{1}\left(\mathcal{B}\right)_{X^{'}}^{Y^{'}},$$
for all $g_{2},h_{2}\in \Omega_{2}\left(\mathcal{B}\right)$ such that $\alpha\left(g_{2}\right) = X^{'}$, $\beta \left( g_{2}\right) = X$, $\alpha \left(h_{2}\right) = Y$, and $\beta \left( h_{2}\right) = Y^{'}$.

Analogously, the material groupoid of the composite is vertically transitive if, and only if,
$$ h_{1}\cdot \Omega_{2}\left(\mathcal{B}\right)_{X}^{Y} \cdot   g_{1}  = \Omega_{2}\left(\mathcal{B}\right)_{X^{'}}^{Y^{'}},$$
for all $g_{1},h_{1}\in \Omega_{1}\left(\mathcal{B}\right)$ such that $\alpha\left(g_{1}\right) = X^{'}$, $\beta \left( g_{1}\right) = X$, $\alpha \left(h_{1}\right) = Y$, and $\beta \left( h_{1}\right) = Y^{'}$. 
\end{proposition}

In particular, if the composite material is horizontally transitive (resp. vertically transitive), all the groups of material symmetries of $\Omega_{1}\left( \mathcal{B}\right)$ (resp. $\Omega_{2}\left( \mathcal{B}\right)$) are conjugated by material isomorphism of the other material, i.e.,
\begin{equation}\label{Matdo5}
g_{2}\cdot \Omega_{1}\left(\mathcal{B}\right)_{X}^{X} \cdot   g_{2}^{-1}  = \Omega_{1}\left(\mathcal{B}\right)_{X^{'}}^{X^{'}} \ \ \left(\text{resp. }g_{1}\cdot \Omega_{2}\left(\mathcal{B}\right)_{X}^{X} \cdot   g_{1}^{-1}  = \Omega_{2}\left(\mathcal{B}\right)_{X^{'}}^{X^{'}}\right),
\end{equation}
for all $g_{2}\in \Omega_{2}\left(\mathcal{B}\right)$ such that $\alpha\left(g_{2}\right) = X $ and $ \beta \left( g_{2}\right) = X^{'} $ (resp. $g_{1}\in \Omega_{1}\left(\mathcal{B}\right)$ such that $\alpha\left(g_{1}\right) = X$ and $\beta \left( g_{1}\right) = X^{'} $).

\begin{proposition}\label{Matdo6}
Consider two uniform materials given by their transitive material groupoids $\Omega_{1}\left( \mathcal{B}\right)$ and $ \Omega_{2}\left( \mathcal{B}\right)$ over the body $\mathcal{B}$. $\boxdot \left( \Omega_{1}\left( \mathcal{B}\right), \Omega_{2}\left( \mathcal{B}\right)\right)$ is horizontally transitive if, and only if, the composite is uniform with $\Omega_{2} \left(\mathcal{B}\right)_{X}^{X} \leq \Omega_{1} \left(\mathcal{B}\right)_{X}^{X}$, for all material particle $X\in \mathcal{B}$.
\begin{proof}
Assume that $\boxdot \left( \Omega_{1}\left( \mathcal{B}\right), \Omega_{2}\left( \mathcal{B}\right)\right)$ is horizontally transitive. Let $X$ be a body point, and $\epsilon \left(X \right)$ its associated identity. Then, for any $g_{2}\in \Omega_{2}\left( \mathcal{B}\right)_{X}$, 
$$g_{2} = g_{2}\cdot \epsilon \left(X \right) \cdot \epsilon \left(X \right)^{-1} \in \Omega_{1}\left( \mathcal{B}\right)$$
i.e., $g_{2}$ is a material isomorphism for both structures of materials. In particular, $\Omega_{2} \left(\mathcal{B}\right)_{X}^{X} \leq \Omega_{1} \left(\mathcal{B}\right)_{X}^{X}$. Conversely, assume that the material composite is uniform and for each material point $X\in \mathcal{B}$, $\Omega_{2} \left(\mathcal{B}\right)_{X}^{X} \leq \Omega_{1} \left(\mathcal{B}\right)_{X}^{X}$. Then, for any two body particles $X, Y \in \mathcal{B}$, there exists a material isomorphism $g \in \Omega_{1}\left( \mathcal{B}\right)_{X}^{Y}\cap \Omega_{2}\left( \mathcal{B}\right)_{X}^{Y}$ and, taking into account that $\Omega_{2} \left(\mathcal{B}\right)_{X}^{X} \leq \Omega_{1} \left(\mathcal{B}\right)_{X}^{X}$, we have that,
$$\Omega_{2} \left(\mathcal{B}\right)_{X}^{Y} = g\cdot \Omega_{2} \left(\mathcal{B}\right)_{X}^{X} \leq g \cdot \Omega_{1} \left(\mathcal{B}\right)_{X}^{X} = \Omega_{1} \left(\mathcal{B}\right)_{X}^{Y}$$
Thus, we have proved that,
$$\Omega_{2} \left(\mathcal{B}\right) \subseteq \Omega_{1} \left(\mathcal{B}\right),$$
Then, for any triple $\left(g_{2},g_{1},h_{2}\right) \in \Omega_{2}\left( \mathcal{B}\right)\times_{ \alpha, \alpha } \Omega_{1}\left( \mathcal{B}\right)\times_{\beta, \alpha} \Omega_{2}\left( \mathcal{B}\right)$, we obtain
\begin{equation*}
 h_{2}\cdot g_{1} \cdot   g_{2}^{-1} \in \Omega_{1}\left(\mathcal{B}\right)
\end{equation*}
\end{proof}
\end{proposition}

Therefore, \textit{horizontal transitivity corresponds to a particular case of uniformity of the composite}; the material symmetries of the material groupoid $\Omega_{2}\left( \mathcal{B}\right)$ are also material symmetries for the other material. We may prove an analogous result for vertical transitivity.

\begin{proposition}\label{dg19}
Consider two uniform materials given by their transitive material groupoids $\Omega_{1}\left( \mathcal{B}\right)$ and $ \Omega_{2}\left( \mathcal{B}\right)$ over the body $\mathcal{B}$. $\boxdot \left( \Omega_{1}\left( \mathcal{B}\right), \Omega_{2}\left( \mathcal{B}\right)\right)$ is vertically transitive if, and only if, the composite is uniform with $\Omega_{1} \left(\mathcal{B}\right)_{X}^{X} \leq \Omega_{2} \left(\mathcal{B}\right)_{X}^{X}$, for all material particle $X\in \mathcal{B}$.
\end{proposition}

\begin{corollary}
$\boxdot \left( \Omega_{1}\left( \mathcal{B}\right), \Omega_{2}\left( \mathcal{B}\right)\right)$ is vertically transitive and horizontally transitive if, and only if, the composite is uniform and $\Omega_{1}\left( \mathcal{B}\right) = \Omega_{2}\left( \mathcal{B}\right)$
\end{corollary}

So, horizontal and vertical transitivity are strictly stronger properties than uniformity of the material composite.\\

\begin{remark}{\rm 
We could intuitively think of these concepts as a uniformity dominated by one of the two materials. For example, in the case of some applications (e.g. metamaterials) one is interested in the dominant character of one of the two materials rather than the character of the other.  }   
\end{remark}

One may consider other, apparently, weaker transitivity properties. Let us consider two uniform materials $\Omega_{1}\left( \mathcal{B}\right)$ and $ \Omega_{2}\left( \mathcal{B}\right)$ over the body manifold $\mathcal{B}$. We will say that the material groupoid of the composite $\boxdot \left( \Omega_{1}\left( \mathcal{B}\right), \Omega_{2}\left( \mathcal{B}\right)\right)$ is \textit{\textbf{weakly horizontally transitive}} if the map $\left(\tilde{\alpha}_{2} , \tilde{\beta}_{2}\right): \boxdot \left( \Omega_{1}\left( \mathcal{B}\right), \Omega_{2}\left( \mathcal{B}\right)\right) \to \Omega_{2}\left( \mathcal{B}\right)\times \Omega_{2}\left( \mathcal{B}\right) $ is surjective. So, the graphical intuition of this condition is as follows,

\begin{equation*} \label{dg16}
\begin{tikzpicture}[baseline=(current  bounding  box.center)]
\draw[thick,red,-stealth'] (0.5,0) -- (0.5,3);
  \draw[thick,red,-stealth'] (-2.5,0) -- (-2.5,3);
\node[above] at (1,1.2) {$g_{2}$};
\node[below] at (-3,1.7) {$h_{2}$};

\node at (2.8,1.5) {\small{Weak hor. tran.}};

\node at (3,1) {$\Longrightarrow$};

  \draw[thick,-stealth'] (8,0) -- (5,0);
  
  \draw[thick,-stealth'] (8,3) -- (5,3);
  \draw[thick,red,-stealth'] (8,0) -- (8,3);
  \draw[thick,,red,-stealth'] (5,0) -- (5,3);
\node at (8.4,1.5) {${g_{2}}$};
\node at (4.6,1.5) {${h_{2}}$};
\node[above] at (6.5,3) {$h_{1}$};
\node[below] at (6.5,-0.1) {$g_{1}$};
\end{tikzpicture}
\end{equation*}

On the other hand, we will say that the material groupoid of the composite $\boxdot \left( \Omega_{1}\left( \mathcal{B}\right), \Omega_{2}\left( \mathcal{B}\right)\right)$ is \textit{\textbf{weakly vertically transitive}} if the map $\left(\tilde{\alpha}_{1} , \tilde{\beta}_{1}\right): \boxdot \left( \Omega_{1}\left( \mathcal{B}\right), \Omega_{2}\left( \mathcal{B}\right)\right) \to \Omega_{1}\left( \mathcal{B}\right)\times \Omega_{1}\left( \mathcal{B}\right) $ is surjective. Graphically:

\begin{equation*} \label{dg18}
\begin{tikzpicture}[baseline=(current  bounding  box.center)]
\draw[thick,-stealth'] (0.5,0) -- (-2.5,0);
  \draw[thick,-stealth'] (0.5,3) -- (-2.5,3);
\node[above] at (-1,3) {$h_{1}$};
\node[below] at (-1,-0.1) {$g_{1}$};

\node at (3,1.5) {\small{Weak ver. tran.}};

\node at (3,1) {$\Longrightarrow$};

  \draw[thick,-stealth'] (8,0) -- (5,0);
  
  \draw[thick,-stealth'] (8,3) -- (5,3);
  \draw[thick,red,-stealth'] (8,0) -- (8,3);
  \draw[thick,,red,-stealth'] (5,0) -- (5,3);
\node at (8.4,1.5) {${g_{2}}$};
\node at (4.6,1.5) {${h_{2}}$};
\node[above] at (6.5,3) {$h_{1}$};
\node[below] at (6.5,-0.1) {$g_{1}$};
\end{tikzpicture}
\end{equation*}

\begin{proposition}
Consider two uniform materials given by their transitive material groupoids $\Omega_{1}\left( \mathcal{B}\right)$ and $ \Omega_{2}\left( \mathcal{B}\right)$ over the body $\mathcal{B}$. $\boxdot \left( \Omega_{1}\left( \mathcal{B}\right), \Omega_{2}\left( \mathcal{B}\right)\right)$ is horizontally (resp. vertically) transitive if, and only if, $\boxdot \left( \Omega_{1}\left( \mathcal{B}\right), \Omega_{2}\left( \mathcal{B}\right)\right)$ it is weakly horizontally (resp. weakly vertically) transitive.

\begin{proof}
We shall focus on the equivalence of the horizontal properties, as the vertical ones are analogous.
The weak horizontal transitivity may be equivalently expressed as follows: for any $g_{2},h_{2}\in \Omega_{2}\left( \mathcal{B}\right)$, there are two arrows $g_{1},h_{1}\in \Omega_{1}\left( \mathcal{B}\right)$, satisfying,
$$ \alpha\left( g_{2}\right) = \alpha\left( g_{1}\right), \ \ \beta\left( g_{2}\right) = \alpha\left( h_{1}\right), \ \ \alpha\left( h_{2}\right) = \beta\left( g_{1}\right), \ \ \beta\left( h_{2}\right) = \beta\left( h_{1}\right),$$
in such a way that $h_{2}\cdot g_{1} = h_{1}\cdot g_{2}$.
In other terms, for any $g_{2},h_{2}\in \Omega_{2}\left( \mathcal{B}\right)$, there exists $g_{1}\in \Omega_{1}\left( \mathcal{B}\right)$, with $\alpha\left( g_{2}\right) = \alpha\left( g_{1}\right)$ and $ \alpha\left( h_{2}\right) = \beta\left( g_{1}\right)$, such that $h_{2}\cdot g_{1} \cdot g_{2}^{-1}\in \Omega_{1}\left( \mathcal{B}\right)$. Taking $g_{2}$ equal to the identity, we have that for any $h_{2}\in \Omega_{2}\left( \mathcal{B}\right)$, there exists $g_{1}\in \Omega_{1}\left( \mathcal{B}\right)$, with $ \alpha\left( h_{2}\right) = \beta\left( g_{1}\right)$, such that $h_{2}\cdot g_{1} \in \Omega_{1}\left( \mathcal{B}\right)$, i.e., $h_{2} \in \Omega_{1}\left( \mathcal{B}\right)$ and, therefore,
\begin{equation}\label{dg20}
\Omega_{2}\left( \mathcal{B}\right) \subseteq \Omega_{1}\left( \mathcal{B}\right)    
\end{equation}
Notice that, \cref{dg20} implies, obviously, that for any $g_{2},h_{2}\in \Omega_{2}\left( \mathcal{B}\right)$, there exists $g_{1}\in \Omega_{1}\left( \mathcal{B}\right)$, with $\alpha\left( g_{2}\right) = \alpha\left( g_{1}\right)$ and $ \alpha\left( h_{2}\right) = \beta\left( g_{1}\right)$, such that $h_{2}\cdot g_{1} \cdot g_{2}^{-1}\in \Omega_{1}\left( \mathcal{B}\right)$. In other words, \cref{dg20} is equivalent to weak horizontal transitivity. Finally, using \cref{dg19} we have proved the result.
\end{proof}
\end{proposition}

Thus, we have proved that two distinct \textit{transitivity conditions} on the double material groupoid give rise to the same type of uniformity.

\subsection*{Strong uniformity}

As we have mentioned in \cref{sec:structure}, it is quite usual to impose the double groupoid to satisfy a \textit{filling condition} in such a way that the double source map $\left(\tilde{\alpha}_{1}, \tilde{\alpha}_{2}\right)$ is a surjective map. There could be some reasons to impose this property; in particular, it is usually useful to guarantee that the domain of multiplication and division maps have the structure of differentiable manifold.\\

Now, we will try to give a physical interpretation of these geometric properties.

\begin{definition}[Strong uniformity]
Let be two uniform materials with material groupoids $\Omega_{1}\left( \mathcal{B}\right)$ and $ \Omega_{2}\left( \mathcal{B}\right)$ over the body $\mathcal{B}$. We will say that the material groupoid of the composite $\boxdot \left( \Omega_{1}\left( \mathcal{B}\right), \Omega_{2}\left( \mathcal{B}\right)\right)$ is \textit{\textbf{strongly uniform}} if the map $\left( \beta \circ \tilde{\beta}_{1}, \tilde{\alpha}_{1}, \tilde{\alpha}_{2}\right) \colon \boxdot \left( \Omega_{1}\left( \mathcal{B}\right), \Omega_{2}\left( \mathcal{B}\right)\right) \to  \mathcal{B} \times_{Id, \beta}\Omega_{1}\left( \mathcal{B}\right)\times_{\left(\alpha, \beta\right), \alpha} \Omega_{2}\left( \mathcal{B}\right) $ is surjective, where $\mathcal{B} \times_{Id, \beta} \Omega_{1}\left( \mathcal{B}\right)\times_{\left(\alpha, \beta\right), \alpha} \Omega_{2}\left( \mathcal{B}\right) $ is given by the triples $\left(X,g_{1},g_{2}\right) \in \mathcal{B}\times \Omega_{1}\left( \mathcal{B}\right) \times \Omega_{2}\left( \mathcal{B}\right)$ such that $X= \beta \left( g_{1}\right)  = \alpha \left( g_{1}\right) = \alpha \left( g_{2}\right)$.
\end{definition}

Equivalently, for any two arrows, a symmetry $g_{1}$ in $\Omega_{1}\left( \mathcal{B}\right)$ and a material isomorphism $g_{2}$ in $\Omega_{2}\left( \mathcal{B}\right)$, we have a  ``\textit{commuting square}'' with these arrows as sides,

\begin{equation*}
\begin{tikzpicture}[baseline=(current  bounding  box.center)]
\fill[black] (-2.6,3) circle (2pt);
\node at (-2.6,3.5) {$X$};

\fill[black] (-2.6,0) circle (2pt);
\node at (-2.6,-0.5) {$X$};

\fill[black] (0.5,0) circle (2pt);
\node at (0.5,-0.5) {$X$};

\draw[thick,-stealth'] (0.3,0) -- (-2.4,0);
  \draw[thick,red,-stealth'] (0.5,0.2) -- (0.5,3);

\node at (1,1.5) {${g_{2}}$};
\node[below] at (-1,-0.1) {$g_{1}$};

\node at (3,1.5) {\small{Strong unif.}};

\node at (3,1) {$\Longrightarrow$};

\fill[black] (5,3) circle (2pt);
\node at (5,3.5) {$X$};

\fill[black] (5,0) circle (2pt);
\node at (5,-0.5) {$X$};

\fill[black] (8,0) circle (2pt);
\node at (8,-0.5) {$X$};

  \draw[thick,-stealth'] (7.8,0) -- (5.2,0);
  
  \draw[thick,-stealth'] (8,3) -- (5.2,3);
  \draw[thick,red,-stealth'] (8,0.2) -- (8,3);
  \draw[thick,red,-stealth'] (5,0) -- (5,2.8);
\node at (8.4,1.5) {${g_{2}}$};
\node at (4.6,1.5) {${h_{2}}$};
\node[above] at (6.5,3) {$h_{1}$};
\node[below] at (6.5,-0.1) {$g_{1}$};
\end{tikzpicture}
\end{equation*}

Roughly speaking, the composite is strongly uniform if for any two materials particles $X$, and $Y$, and two material isomorphisms $j_{X,X}^{1}\phi_{1} \in \Omega_{1}\left( \mathcal{B}\right)$ and $j_{X,Y}^{1}\phi_{2} \in \Omega_{2}\left( \mathcal{B}\right)$, there exists another two material isomorphisms $j_{Y,X}^{1}\psi_{1} \in \Omega_{1}\left( \mathcal{B}\right)$ and  $j_{X,X}^{1}\psi_{2} \in \Omega_{2}\left( \mathcal{B}\right)$ in such a way that
$$j_{X,X}^{1}\psi_{2}\circ j_{X,X}^{1}\phi_{1} = j_{Y,X}^{1}\psi_{1} \circ j_{X,Y}^{1}\phi_{2}.$$
In other terms, any two points may be connected composing a material symmetry of the first material ($\Omega_{1}\left( \mathcal{B}\right)$) at $X$ with a material isomorphism of the second material ($\Omega_{2}\left( \mathcal{B}\right)$) from $X$ to $Y$, or equivalently, composing a material symmetry of the second material ($\Omega_{2}\left( \mathcal{B}\right)$) at $X$ with a material isomorphism of the first material ($\Omega_{1}\left( \mathcal{B}\right)$) from $X$ to $Y$. In other words, it satisfies a certain type of ``\textit{commuting}'' condition between the two material structures.

\begin{proposition}
Let be two uniform materials with material groupoids $\Omega_{1}\left( \mathcal{B}\right)$ and $ \Omega_{2}\left( \mathcal{B}\right)$ over the body $\mathcal{B}$. If $\boxdot \left( \Omega_{1}\left( \mathcal{B}\right), \Omega_{2}\left( \mathcal{B}\right)\right)$ satisfies the strong uniformity, the material composite is uniform.
\begin{proof}
Let be $X$ and $Y$ two body points, $\epsilon \left(X \right)$ the identity at $X$, and $g_{2}\in  \Omega_{2}\left( \mathcal{B}\right)_{X}^{Y}$. Then, for the strong uniformity, there exist $h_{2}\in \Omega_{2}\left( \mathcal{B}\right)_{X}^{X}$ and $h_{1}\in \Omega_{1}\left( \mathcal{B}\right)_{Y}^{X}$, such that,
$$h_{2} = h_{2} \cdot \epsilon \left(X \right)=  h_{1}\cdot g_{2} $$
\begin{equation*}
\begin{tikzpicture}[baseline=(current  bounding  box.center)]
\draw[thick] (0.5,0) -- (-2.5,0);
\draw[thick] (0.5,-0.1) -- (-2.5,-0.1);

\draw[thick,red,-stealth'] (0.5,0) -- (0.5,3);

\node at (1,1.5) {${g_{2}}$};
\node[below] at (-1,-0.1) {$\epsilon \left( X\right)$};

\node at (3,1) {$\Longrightarrow$};

  \draw[thick] (8,0) -- (5,0);
  \draw[thick] (8,-0.1) -- (5,-0.1);
  
  \draw[thick,-stealth'] (8,3) -- (5,3);
  \draw[thick,red,-stealth'] (8,0) -- (8,3);
  \draw[thick,,red,-stealth'] (5,0) -- (5,3);
\node at (8.4,1.5) {${g_{2}}$};
\node at (4.6,1.5) {${h_{2}}$};
\node[above] at (6.5,3) {$h_{1}$};
\node[below] at (6.5,-0.1) {$\epsilon \left( X\right)$};
\end{tikzpicture}
\end{equation*}

Hence, $h_{1}=h_{2}\cdot g_{2}^{-1}\in\Omega_{2}\left( \mathcal{B}\right)$ is a material isomorphism from $Y$ to $X$ for both materials. Therefore, the composite is uniform.

\end{proof}

\end{proposition}

In this manner, we prove that the transitivity property constitutes a stricter condition than the uniformity of the composite. Specifically, we not only impose the requirement that the composite possesses common material isomorphisms—ensuring its uniformity—but also demand the existence of a certain form of compatibility between the two material structures. This compatibility can be characterized by a specific type of commutativity between the material isomorphisms associated with both mechanical responses.\\

\begin{example}[Crystalline solids]
Let us present an example of a uniform material that does not satisfy the conditions of a strongly uniform material. Let $\mathcal{B}$ be a \textit{solid}, and $\varphi_{0} \colon \mathcal{B} \rightarrow \mathbb{R}^{3}$ an undistorted reference configuration. Then, for any body point $X \in \mathcal{B}$, and each material symmetry $j_{X,X}^{1}\phi$, the associated matrix to $j_{\varphi_{0}\left(X\right), \varphi_{0}\left(X\right)}^{1}\varphi_{0}\circ \phi \circ \varphi_{0}^{-1},$
is a orthogonal matrix. A particular case is that of the so-called \textit{crystalline solids}, which are classified into thirty-two classes. Each class is characterized by a specific type of symmetry group, such that these symmetry groups are finite subgroups of the orthogonal group (\cite{COLE2,MEPMSoCR}).

Thus, we will consider two uniform crystalline solids, whose material material groupoids are given by  $\Omega_{1}\left( \mathcal{B}\right)$ and  $\Omega_{2}\left( \mathcal{B}\right)$. In particular, the material groupoid $\Omega_{1}\left( \mathcal{B}\right)$ satisfies that for each material particle $X \in \mathcal{B}$, there exist only three material symmetries at $X$, $\epsilon \left( X \right), g_{X}, g_{X}^{-1}$, so that $g_{X}$ is not trivial. In other words, the material symmetries at $X$ are characterized by $g_{X}$. Furthermore, for any two body point $X$ and $Y$, let us highlight a material isomorphism $g_{X,Y}$ from $X$ to $Y$ (in case $X=Y$, we fix $g_{X,Y} =\epsilon\left( X \right) $). Then, all the material isomorphisms from $X$ to $Y$ are of the form $g_{X,Y}$, $g_{X,Y} \cdot g_{X}$ and $g_{X,Y} \cdot g_{X}^{-1}$.

Assume now that the second uniform crystalline solid, whose material groupoid is $\Omega_{2}\left( \mathcal{B}\right)$, fullfils the same property, i.e., for any material point $X \in \mathcal{B}$, there exist only three material symmetries at $X$, $\epsilon \left( X \right), h_{X}, h_{X}^{-1}$, so that $h_{X}$ is not trivial and different to $g_{X}$. In addition, for any two body points $X$ and $Y$, $g_{X,Y}$ is a material isomorphism from $X$ to $Y$. Then, all the material isomorphisms from $X$ to $Y$ are of the form $g_{X,Y}$, $g_{X,Y} \cdot h_{X}$ and $g_{X,Y} \cdot h_{X}^{-1}$.

Then, for any $X$ and $Y$, both groupoids share a material isomorphism, and as a consequence, the material composite is uniform. Let us consider the arrows $g_{X}\in \Omega_{1}\left( \mathcal{B}\right)$ and $g_{X,Y}\cdot h \in \Omega_{2}\left( \mathcal{B}\right)$. We then ask whether the following square can be completed,

\begin{equation*}
\begin{tikzpicture}[baseline=(current  bounding  box.center)]
\fill[black] (-2.6,3) circle (2pt);
\node at (-2.6,3.5) {$X$};

\fill[black] (-2.6,0) circle (2pt);
\node at (-2.6,-0.5) {$X$};

\fill[black] (0.5,0) circle (2pt);
\node at (0.5,-0.5) {$X$};

\draw[thick,-stealth'] (0.3,0) -- (-2.4,0);
  \draw[thick,red,-stealth'] (0.5,0.2) -- (0.5,3);

\node at (1.4,1.5) {${g_{X,Y}\cdot h_{X}}$};
\node[below] at (-1,-0.1) {$g_{X}$};

\end{tikzpicture}
\end{equation*}

In other terms, are there $G \in \Omega_{1}\left( \mathcal{B}\right)$ and $H \in \Omega_{2}\left( \mathcal{B}\right)$ such that
$$H^{-1}G = g_{X,Y}\cdot h_{X}\cdot g_{X}^{-1} \text{?}$$
By considering all possible cases, this occurs only if one of the following equalities holds:
\begin{enumerate}[i)]
    \item $g_{X,Y}\cdot h_{X}^{-1} = g_{X,Y}\cdot h_{X}\cdot g_{X}^{-1}$
     \item $g_{X,Y}\cdot g_{X}\cdot h_{X}^{-1} = g_{X,Y}\cdot h_{X}\cdot g_{X}^{-1}$
     \item $g_{X,Y}\cdot g_{X}^{-1}\cdot h_{X}^{-1} = g_{X,Y}\cdot h_{X}\cdot g_{X}^{-1}$
     \item $g_{X,Y}\cdot h_{X} = g_{X,Y}\cdot h_{X}\cdot g_{X}^{-1}$
     \item $g_{X,Y}\cdot g_{X}\cdot h_{X} = g_{X,Y}\cdot h_{X}\cdot g_{X}^{-1}$
     \item $g_{X,Y}\cdot g_{X}^{-1}\cdot h_{X}= g_{X,Y}\cdot h_{X}\cdot g_{X}^{-1}$
\end{enumerate}
\end{example}

Notice that $i)$ is equivalent to $ h_{X}^{2} = g_{X}^{2}$, which is absurd by construction of both groupoids. Analogously, $iv)$ is equivalent to $g_{x} = \epsilon \left( X \right)$ which is, again, absurd.

On the other hand, the rest of the equalities may be equivalently rewritten as follows,

\begin{itemize}
     \item[ii)] $g_{X}\cdot h_{X}^{-1} \cdot g_{X} =  h_{X}$
     \item[iii)] $g^{-1}_{X}\cdot h_{X}^{-1} \cdot g_{X} =  h_{X}$
     \item[v)] $g_{X}\cdot h_{X} \cdot g_{X} =  h_{X}$
     \item[vi)] $g^{-1}_{X}\cdot h_{X} \cdot g_{X} =  h_{X}$
\end{itemize}

In other words, the material symmetries must satisfy one of the following four conjugacy conditions between them. Therefore, we only have to formulate a particular example in which the symmetries does not satisfy these identities.

Observe that this provides a clear example to intuitively grasp the distinction between uniformity and strong uniformity. In the latter case, the material isomorphisms of both material structures must satisfy an additional compatibility condition between them, translated into these four conjugacy equalities.

\subsection*{Weak uniformity}
Let us now consider another property of ``\textit{completing squares}'', which (in this case) will be less restrictive than uniformity of the composite.
\begin{definition}[Weak uniformity]
Let be two uniform materials given by their transitive material groupoids $\Omega_{1}\left( \mathcal{B}\right)$ and $ \Omega_{2}\left( \mathcal{B}\right)$ over the body manifold $\mathcal{B}$. We will say that the material groupoid of the composite $\boxdot \left( \Omega_{1}\left( \mathcal{B}\right), \Omega_{2}\left( \mathcal{B}\right)\right)$ is \textit{\textbf{weak uniform}} if the map $\left(\alpha \circ \tilde{\alpha}_{1}, \beta \circ \tilde{\beta}_{1}, \alpha \circ \tilde{\alpha}_{2}, \beta \circ \tilde{\beta}_{2}\right): \boxdot \left( \Omega_{1}\left( \mathcal{B}\right), \Omega_{2}\left( \mathcal{B}\right)\right) \to \mathcal{B} \times \mathcal{B}\times \mathcal{B} \times \mathcal{B}$ is surjective.
\end{definition}
Roughly speaking, the composite material groupoid is weak uniform if for any four material particles $X,Y,Z,$ and $T$, there exists a square in the composite material groupoid whose \textit{corners} are these points, namely \\

\begin{equation*} \label{dg17}
\begin{tikzpicture}[baseline=(current  bounding  box.center)]
\fill[black] (0.5,0) circle (2pt);
\fill[black] (-2.5,0) circle (2pt);
\fill[black] (0.5,3) circle (2pt);
\fill[black] (-2.5,3) circle (2pt);

\node at (3,1.5) {\small{Weak unif.}};

\node at (3,1) {$\Longrightarrow$};

\fill[black] (8,0) circle (2pt);
\fill[black] (5,0) circle (2pt);
\fill[black] (5,3) circle (2pt);
\fill[black] (8,3) circle (2pt);

  \draw[thick,-stealth'] (7.9,0) -- (5.1,0);
  \draw[thick,-stealth'] (7.9,3) -- (5.1,3);
  \draw[thick,red,-stealth'] (8,0.1) -- (8,2.9);
  \draw[thick,,red,-stealth'] (5,0.1) -- (5,2.9);
\node at (8.4,1.5) {${g_{2}}$};
\node at (4.6,1.5) {${h_{2}}$};
\node[above] at (6.5,3) {$h_{1}$};
\node[below] at (6.5,-0.1) {$g_{1}$};
\end{tikzpicture}
\end{equation*}

Thus, a material composite satisfies this property if any four material points may be \textit{connected} in two different ways, by two material isomorphisms, one of each material body. More precisely, for any choice of four material points $X,Y, Z, T \in \mathcal{B}$, any two of them, let say $X$ and $Y$, may connected by a $1-$jet $j_{X,Y}^{1}\phi$ of a local diffeomorphism $\phi$, such that may be decomposed into a composition of a material isomorphism of the first material, $j_{X,Z}^{1}\phi_{1} \in \Omega_{1}\left( \mathcal{B}\right)$, from $X$ to some $Z$, with a material isomorphism of the second material, $j_{Z,Y}^{1}\phi_{2} \in \Omega_{2}\left( \mathcal{B}\right)$, from $Z$ to $Y$, or equivalently, into a composition of a material isomorphism of the second material, $j_{X,T}^{1}\psi_{2} \in \Omega_{2}\left( \mathcal{B}\right)$, from $X$ to another material point $T$, with a material isomorphism of the first material, $j_{T,Y}^{1}\psi_{1} \in \Omega_{1}\left( \mathcal{B}\right)$, from $T$ to $Y$, i.e.,
\begin{equation}
  j_{X,Y}^{1}\phi = j_{Z,Y}^{1}\phi_{2} \cdot j_{X,Z}^{1}\phi_{1} = j_{T,Y}^{1}\psi_{1} \cdot j_{X,T}^{1}\psi_{2}  
\end{equation}

In other words, it satisfies a certain type of ``\textit{commuting}'' condition between the two material structures, allowing us to intuit that there is a certain \textit{compatibility} between the material properties of the two materials, one that is more general than uniformity.

\begin{proposition}
Consider two uniform materials, with transitive material groupoids $\Omega_{1}\left( \mathcal{B}\right)$ and $ \Omega_{2}\left( \mathcal{B}\right)$, in such a way that the composite is uniform. Then, $\boxdot \left( \Omega_{1}\left( \mathcal{B}\right), \Omega_{2}\left( \mathcal{B}\right)\right)$ is weak uniform.
\begin{proof}
Let be $X,Y,Z,$ and $T$ four material points. Then, by the uniformity of the composite, there exist three material isomorphisms for both materials $g_{1},g_{2},h_{2}$ with $\alpha\left(g_{1}\right) = X = \alpha\left(g_{2}\right) $, $\beta\left(g_{1}\right) = Y = \alpha\left(h_{2}\right) $, $\beta\left(g_{2}\right) = T$, and $\beta\left(h_{2}\right) = Z$. So, the square $\left(\left(g_{1},h_{2}\cdot g_{1}\cdot g_{2}^{-1}\right), \left(g_{2},h_{2}\right)\right) \in \boxdot \left( \Omega_{1}\left( \mathcal{B}\right), \Omega_{2}\left( \mathcal{B}\right)\right)$, completes the square with corners $X,Y,Z,$ and $T$.

\begin{equation*}
\begin{tikzpicture}[baseline=(current  bounding  box.center)]
  \draw[thick,-stealth'] (3,0) -- (0,0);
  \draw[thick,-stealth'] (3,3) -- (0,3);
  \draw[thick,red,-stealth'] (3,0) -- (3,3);
  \draw[thick,red,-stealth'] (0,0) -- (0,3);
\node at (3.5,1.5) {${g_{2}}$};
\node at (-0.5,1.5) {${h_{2}}$};
\node[above] at (1.5,3) {$h_{2}\cdot g_{1}\cdot g_{2}^{-1}$};
\node[below] at (1.5,0) {$g_{1}$};
\end{tikzpicture}
\end{equation*}
\end{proof}

\end{proposition}

In this way, weak uniformity is a property which satisfies any uniform composite. However, the converse is not true. Let us give an example.

\begin{example}[Triclinic crystal]
Another particular example of solid crystal are those called \textit{triclinic crytals}. Following \cite{MEPMSoCR}, it is a solid crystal with the trivial symmetry groups (there are no symmetries other than the identities).

Now, consider two uniform triclinic crystals whose material groupoids are characterized by $\Omega_{1}\left( \mathcal{B}\right)$ and $\Omega_{2}\left( \mathcal{B}\right)$. Furthermore, the material isomorphisms are determined by two different implants $P_{i}: \mathcal{B} \to \Omega_{i}\left( \mathcal{B}\right)^{T}$, with $T\in \mathcal{B}$, which are sections of the respective source maps, in such a way, for all $X,Y\in \mathcal{B}$,
$$P_{i}\left( Y \right)^{-1}\cdot P_{i}\left( X \right) \in \Omega_{i}\left( \mathcal{B}\right)_{X}^{Y}$$
Finally, we will impose that the material isomorphisms of the different materials commutes, i.e.,
$$P_{2}\left( Z \right)^{-1}\cdot P_{2}\left( Y \right) \cdot P_{1}\left( Y \right)^{-1}\cdot P_{1}\left( X \right) = P_{1}\left( Z \right)^{-1}\cdot P_{1}\left( Y \right) \cdot P_{2}\left( Y \right)^{-1}\cdot P_{2}\left( X \right)$$
Then, the composite is weak transitive but it is \textit{completely non uniform}: there are no material isomorphisms between whatever different material particles.
\end{example}

This example may help us intuitively grasp the distinction between weak uniformity and uniformity. Additionally, it illustrates the type of compatibility that guarantees the weak uniformity of the material composite.


\section{Conclusions and further work}

The abstract structure of double groupoid has permitted us to introduce several notions of uniformity for composites, in particular the notions of horizontal and vertical uniformity, and weak uniformity using the method of \textit{completing squares}. We have also studied the relations between these different notions.

As a future work, we are working in the following items.

\begin{itemize}
    
\item To extend the concepts of uniformity to composites of an arbitrary number of materials.

\item To extend these notions to materials with microstructure and liquid crystals. 

\item To study the homogeneity of uniform composite materials using the concept of double algebroid, as well as to reinterpret the results in terms of $G$-structures.

\item To study the case of smooth uniformity.

\end{itemize}
\section*{Acknowledgments}
M. de Le\'on and V. M. Jiménez received financial support from Grant PID2022-137909NB-
C21 funded by MICIU/AEI/10.13039/501100011033. M. de León also acknowledges financial support from Grant
CEX2023-001347-S funded by MICIU/AEI/10.13039/501100011033.
V. M. Jiménez acknowledge the finantial support from MICIU and European Union-\textit{NextGenerationUE}. \\


\bibliographystyle{plain}

\bibliography{Library}

\end{document}